\documentclass[fleqn,usenatbib]{mnras}

\usepackage{newtxtext,newtxmath}
\usepackage{soul}
\usepackage{float}
\usepackage[T1]{fontenc}
\DeclareRobustCommand{\VAN}[3]{#2}
\let\VANthebibliography\thebibliography
\def\thebibliography{\DeclareRobustCommand{\VAN}[3]{##3}\VANthebibliography}

\providecommand{\zp}[1]{{\textbf{\textcolor{purple}{[zephyr: #1]}}}}
\providecommand{\bs}[1]
{{\textbf{\textcolor{orange}{[Bianca: #1]}}}}
\usepackage[dvipsnames]{xcolor}
\usepackage{multirow}
\usepackage{array}
\usepackage{booktabs}

\usepackage{graphicx}
\usepackage{subcaption}
\usepackage{amsmath}
\usepackage{amssymb}
\usepackage{xcolor}
\usepackage{changes}
\usepackage{adjustbox}
\usepackage{hyperref}
\usepackage{comment}

\title[Binaries' encounters with SgrA*]
{Dynamics of recaptures, ejections and mergers of stellar mass binaries over multiple encounters with SgrA*}
\author[B. Sersante et al.]{
Biancamaria Sersante,$^{1}$\thanks{E-mail: sersante@strw.leidenuniv.nl}
Zephyr Penoyre,$^{1}$
Elena Maria Rossi $^{1}$
\\
$^{1}$ Leiden Observatory, Leiden University, PO Box 9513, 2300 RA Leiden,
The Netherlands}

\date{Received XX. Accepted XX}

\pubyear{2023}

\begin{document}
\label{firstpage}
\pagerange{\pageref{firstpage}--\pageref{lastpage}}
\maketitle


\begin{abstract}
A common origin for a host of stellar phenomena in galactic centres is the tidal encounter between stellar binaries and a massive black hole (MBH), known as the ``Hills mechanism''. Following the encounter, binaries may disrupt into an ejected star and a captured one, they may merge, or survive to either \textit{fly away} or \textit{come back} for one or more subsequent encounters, until they are either disrupted or fly-away. In this paper, we analyse how a binary’s fate depends on its orbital parameters, by following its evolution through up to three subsequent pericentre passages. We choose an initial population of circular binaries on parabolic orbits. We present results from our restricted three-body formalism, whose strength lies in the ability to easily explore a multidimensional parameter space and make predictions independent of the binary physical properties. We find that fates depend strongly on orbital inclination, how deep the encounter is into the MBH tidal sphere and on the binary eccentricity, developed during encounters. Generally, non-retrograde trajectories, high eccentricities or deep encounters produce disruptions preferentially. Disruption is the most common fate. A significant fraction of the surviving binaries fly away at velocities typically two orders of magnitude smaller than those of ejected stars. Multiple encounters boost disruptions by $20\%$ or more. Finally, using an example system, we investigate the effect of finite stellar sizes and lifetimes, showing that mergers occur $31\%$ of the time, and that disruptions are still boosted by $\sim 10\%$ through subsequent passages.

\end{abstract}



\section{Introduction}
\label{sec:intro}

Much has been discovered about the Galactic Centre (GC) of the Milky Way (MW). Observational milestones include Jansky’s radio detection of our Massive Black Hole (MBH), SgrA*, in 1931 \citep[see][]{Jansky}, the development of infrared astronomy \citep[e.g.][]{Becklin}, to track trajectories of stars near Sgr A* in the early 2000s \citep[see e.g.][]{Schodel,Ghez2003,Ghez2005}, and the first image of Sgr A*, released by \citep[][]{EHT2022} . Nevertheless, many questions remain unanswered, the origin and assembly history of Sgr A*, of the GC and, in particular, of its stellar populations. The GC complex stellar dynamics is a rich field of study too, tightly linked to the largely unknown rates of many high-energy transients, including Tidal Disruption Events (TDEs) and gravitational wave sources where stellar mass black holes spiral inwards torwards the central MBH, called Extreme Mass Ratio Inspirals (EMRIs).

Since direct observation of the GC is challenged by obscuration and stellar crowding, a complementary and captivating tool to explore these questions is hypervelocity stars (HVSs). These are stars ejected from the GC at speeds up to a few thousands of km s$^{-1}$, high enough to be observable in the halo on unbound trajectories from the Galaxy. HVSs can be used as tracers, as they carry information about their native GC to regions that are more easily observationally accessible. Currently, only one candidate has been successfully traced back to the GC and thus confirmed as an HVS by \citep[][]{Koposov}: S5-HVS1, an A-type main sequence star, with a velocity of  $ 1755\pm 50$km s$^{-1}$.  The number of promising candidates is around a dozen \citep[see][]{Brown2014, BrownLattanzi18, BromleyKenyon18}. Various methods have been suggested to improve observations 
\citep[see e.g.][]{KenyonHVSLMC,Marchetti2022,Evans2022,Evans2022II,Evans2023,Sill2024} and several mechanisms have been proposed to explain the origin of such fast stars.
For instance, the ejection of an HVS could be the result of the close interaction between a globular cluster and a supermassive black hole \citep[see e.g.][]{Capuzzo2015,Fragione2016,Fragione2017}, or of the three-body interaction between a star and a binary black hole
\citep[see e.g.][]{Yu2003,Sesana2007,GinsburgLoeb2007,Sesana2008,Marchetti2018,Rasskazov2019}. In this paper, we focus on another possible explanation which relies on the Hills mechanism, namely the tidal separation of a binary stellar system by SgrA*, which may result in a binary component being ejected while the other remains bound to the MBH on a tight eccentric orbit \citep[see e.g.][]{Hills1988,Sari10,Kobayashi12,Rossi2014,Brown18}. 

\citet{Verberne2025} showed that this mechanism can simultaneously explain S5-HVS1 and the presence of the young stellar cluster around SgrA* called the S-star cluster \citep[see for instance][]{Ghez2008,Gillessen2009}. In this scenario, S-stars are the previous binary companions of HVSs. In addition, the Hills mechanism has been invoked as a dynamical channel to create EMRIs \citep{miller05,raveh21}. In particular, \cite{LinialSari} 
found that it is expected to contribute to GW driven stellar EMRIs for galaxies with $M\gtrsim 10^5M_{\odot}$, which encompasses SgRA$^{*}$.
This mechanism may also be responsible for at least a subset of Quasi-Period Eruptions (QPEs) observed in X-rays \citep[e.g.][]{LinialSari}. \cite{SariFragione} showed that the Hills disruption of stellar binaries in the vicinity of a SMBH may affect the shape of the density stellar cusp, which in turn affects the rate of stellar disruption (TDEs) and EMRIs. These are among the scientific motivations behind this paper's unprecedented detailed dynamical analysis of the Hills mechanism.

Briefly, the characteristic scales of Hills mechanism ejecta can be described as follows. Given a binary with semi-major axis $a_{\rm b}$ and total mass $m$ interacting on a parabolic trajectory with a MBH with mass $M$, the tidal forces of the latter overcome the self-gravity of the binary at an approximate distance $r_{\rm t} \approx a_{\rm b}(M/m)^{1/3}$ from the MBH, called the {\it tidal separation radius} or in short, {\it tidal radius}. Once the binary crosses the tidal radius, it may be tidally separated, resulting in one of the binary members (e.g. $m_{\rm ej}$) being ejected and the other (e.g. $m_{\rm cap}$) captured by the MBH. 
The ejected star will have characteristic ejection velocity of the order of
\begin{align*}
 \nu &=\left(\frac{M}{m}\right)^{\frac{1}{6}}\sqrt{\frac{2Gm_{\rm cap}}{a_{\rm b}}}\\
 &\approx 1300 \; \mathrm{km}\; \mathrm{s}^{-1}\left(\frac{M}{4\times 10^6M_{\odot}}\right)^{\frac{1}{6}}\left(\frac{m}{4M_{\odot}}\right)^{-\frac{1}{6}}\left(\frac{m_{\rm cap}}{3M_{\odot}}\right)^{\frac{1}{2}}\left(\frac{a_{\rm b}}{0.1 \mathrm{AU}}\right)^{-\frac{1}{2}}
\end{align*}
and the captured component will end up on a bound orbit with semi-major axis of order 
\begin{align*}
    \alpha &=\frac{1}{2}Q^{\frac{2}{3}}\frac{m}{m_{\rm ej}}a_{\rm b}\\
    &\approx 3\cdot 10^{-3} \ \mathrm{pc} \left(\frac{M}{4\times 10^6M_{\odot}}\right)^{\frac{2}{3}}\left(\frac{m}{4M_{\odot}}\right)^{\frac{1}{3}}\left(\frac{m_{\rm ej}}{M_{\odot}}\right)^{-1}\left(\frac{a_{\rm b}}{0.1 \mathrm{AU}}\right),
\end{align*}
potentially ejecting one star from the Milky Way and leaving the other on a very close orbit to SgrA*.

This, however, is not the only possible outcome of this dynamical encounter.

In their work, \citep[][]{Sari10} found that there is a non-null probability that a binary survives disruption. In this case, there are two possible outcomes: binary members can merge (we call this channel {\it Ms} for ``mergers'') or they can remain bound to each other. Mergers as a possible explanation of G-type objects \citep[see e.g.][]{Ciurlo,Campbell,chu,Jia}, have been analysed by \cite{StephanNaoz2016,StephanNaoz} through the Eccentric Kozai-Lidov Mechanism and accounting for stellar evolution for binaries outside the BH tidal radius. With our work, we can complement these results by considering mergers that are purely dynamical in nature, in a non-perturbative and non-secular regime, encompassing also mergers of binaries diving into the tidal sphere through multiple encounters.
Mergers have also been analysed by \cite{MandelLevin2015} and \cite{BradnickMandelLevin2017} for a population of eccentric binaries with specific distributions of their orbital parameters. The authors defined as ``mergers'' the cases when the distance between the two stars becomes smaller than the sum of their radii (as in \cite{Sari10}) and integrated the evolution of the system using REBOUND \citep[][]{ReinLiu2012,ReinSpiegel2015}. Our semi-analytical approach can be used as an independent and complementary analysis of mergers after one encounter and provides new information on multiple passages between pericentres. 

On the other hand, if they remain bound, binaries can either end up on a bound orbit around the MBH and thus come back to interact with it again in a second gravitational encounter, or they can end up on unbound trajectories and fly away to populate the GC. In this paper, we call these two different types of binaries coming-back (CBs) and flying-away (FAs), respectively. CBs can then either survive the second encounter with the MBH (as a CB or a FA binary) or they can be disrupted and dissolve into a HVS and an S-star (we dub disrupted binaries as Ds). This series of events can occur for multiple subsequent pericentre passages of their centre-of-mass (CM) orbit.  

The dynamical interactions of stellar binaries with an MBH have previously been investigated with different methods: three-body scattering experiments \citep[][]{Sesana2007, BromleyKenion2006,GenerozovPerets2022}, full N-body simulation of a galactic nucleus \citep[e.g.][]{AntoniniFaber2010,AntoniniLombardi2011,ProdanAntonini2015,MandelLevin2015,BradnickMandelLevin2017}, and with the restricted three-body framework \citep[e.g.][]{Sari10, Kobayashi12, Brown18}. The latter takes advantage of the extreme mass ratio between the binary and the BH to linearise the equation of motion and energy, so that they can provide accurate results\footnote{We tested the accuracy of the approximation against full 3-body simulations (similarly to what was done by \cite{Sari10})} with less computational resources; remarkably, these results depend only on the geometry of the encounter and eccentricities, but not on the physical properties of the binary. 

In this paper, we use the restricted three-body framework, extending this formalism so as to be able to follow multiple encounters for the first time. In particular, our goal is to provide the following.
\begin{enumerate}
    \item An identification of the orbital parameter space that mostly contributes to any given outcome.
    \item An assessment of how the single encounter distribution of ejection velocities is affected by a second- and third-generation of ejected stars, and of the fraction of those that can be considered as HVSs (in this work we define them as stars ejected from the GC with velocity in excess of 1000 km s$^{-1}$).
    \item A description of the properties of S-stars and FAs.
    \item The fraction of binary mergers.
\end{enumerate}
The main novelty with respect to previous papers consists in the detailed dynamical description of the outcomes (ratio and properties) of multiple pericentre passages, where our results are independent of the binary physical properties, such as masses, mass ratios and initial separation. Additionally, we give an unprecedented description of FAs that, as far as we know, have only been previously mentioned but not analysed by \citet{MandelLevin2015}.

This paper is organized as follows. In Section \ref{P1}, we introduce the restricted three-body problem, by detailing how to compute the binary and CM-orbit properties. In Section \ref{distr} and \ref{Multiplep} we present results after one and three pericentre passages, respectively. In Section \ref{Results}, we present the distributions of binary and CM-orbit properties (for ejected and captured stars and for FA binaries) in a set of units rescaled with respect to the initial binary semi-major axis, and normalized with respect to the initial binary energy and angular momentum. These choices guarantee that our results hold for a generic binary. Additionally, using physical units, we provide an estimate on the predicted ratio of HVSs. Finally, in Section \ref{Concl} we discuss these results and draw our conclusions. 


\vspace{-0.5em}
\section{Restricted three-body problem }\label{3bp}
\label{P1}

The general three body problem is significantly simplified if we restrict ourselves to an encounter between a stellar-mass binary (with total $m$) and a Massive Black Hole (MBH) (of mass $M$). We can assume that the binary components are initially much closer to each other than the massive object, and that $Q \equiv M/m$ is $Q\gg 1$. For example when we consider stellar-mass binaries orbiting SgrA* ($M\simeq 4 \times 10^{6}M_\odot$) then $Q$ is of order $10^{6}$. 
This allows for the approximation of a ``restricted three-body problem''\footnote{Typically, the restricted three-body formalism is applied to a system where only one of the three bodies can be considered as a test mass, e.g. the Moon in the Earth-Sun-Moon case. However, both the typical case and ours are particular cases of ``reductions'' of the three-body problem which assume that some gravitational terms can be ignored, i.e. that some of the smaller masses contribute negligibly to the dynamics of the larger masses.} where the MBH is taken to be always stationary. Following \citet{Sari10,Kobayashi12} we solve the motion of the binary Center of Mass (CM) around the MBH a priori, as a simple keplerian orbit. We then integrate the evolution of the binary system as its CM follows that fixed trajectory. 

We now generically label the binary members $1$ and $2$ such that $m=m_1+m_2$ and define the binary mass ratio $q=\frac{m_2}{m_1}$ (which we take to be $\leq 1$ always). Practically, we set out to calculate, as a function of time, the distance of each binary member from the MBH, $\mathbf{r}_{1} = \mathbf{r}_{\rm cm} - (m_{2}/m) \mathbf{r}_{\rm b}$ and $\mathbf{r}_{2} = \mathbf{r}_{\rm cm} + (m_{1}/m) \mathbf{r}_{\rm b}$, where $\mathbf{r}_{\rm cm}$ is the CM distance, and $\mathbf{r}_{\rm b} \equiv \mathbf{r}_{2}-\mathbf{r}_{1}$ the binary separation.
We define the tidal radius (i.e. the characteristic distance from the MBH at which a binary would be expected to separate) as $r_{\rm t} = Q^\frac{1}{3} a_{\rm b}$, where $a_{\rm b}$ is the initial binary semi-major axis. Now we can make more explicit the assumption that the binary separation is initially relatively small: we require $a_{\rm b} \ll r_{\rm t}$ initially.
If above conditions are satisfied, the formalism we present could apply to any scale, including planetary systems and asteroids, and any type of $\sim$ point mass object, including compact stellar remnants like black holes and neutron stars. However, given the case of interest here, we will generally refer to the massive object as a MBH and to the binary components as stars.

Throughout this work we will refer to the motion of the binary's CM around the MBH as \emph{CM trajectory}, denoted with the subscript \emph{cm}. On the other hand, we call the orbit of the two stars the \emph{binary's orbit} and denote their properties and characteristics with the subscript \emph{b}.

\subsection{The Centre of Mass's trajectory}
\label{sec:CM}
In our restricted three-body problem the MBH is always stationary at the origin of our coordinate system. During each passage the CM moves along a fixed trajectory. Although we start the first encounter on a parabolic orbit, the trajectory changes between successive passages and therefore the CM trajectory is a generic conic orbit described by closest approach distance (pericentre) $r_{\rm p}$ and eccentricity $e_{\rm cm}$ and has position
\begin{equation}
 \mathbf{r}_{\mathrm{{\mathrm{cm}}}}=\frac{r_{\mathrm{p}}(1+e_{\mathrm{{\mathrm{cm}}}})}{1+e_{\mathrm{{\mathrm{cm}}}}\cos f}\begin{bmatrix}
\cos f \\
\sin f\\
0
\end{bmatrix},
\label{eq:cm_vector}
\end{equation}
and velocity 
\begin{equation}
\mathbf{v}_{\mathrm{{\mathrm{cm}}}}=\sqrt{\frac{GM}{r_{\rm p}(1+e_{\mathrm{{\mathrm{cm}}}})}}
    \begin{bmatrix}
-\sin f  \\
e_{\mathrm{{\mathrm{cm}}}}+\cos f\\
0
\end{bmatrix}.\\
\end{equation}
Here we have chosen the orientation of our coordinate system such that the CM trajectory is confined to the $I-J$ plane, with the CM passing through the $I$-axis at periapse (see Fig. \ref{fig:cartoon_angles}). The true anomaly, $f$, is the phase of the CM trajectory (with $f=0$ at periapse) and follows
\begin{equation}\label{dotf}
\dot{f}=\sqrt{\frac{GM}{r_{\rm p}^3}}(1+e_{\mathrm{{\mathrm{cm}}}})^{-3 / 2}(1+e_{\mathrm{{\mathrm{cm}}}} \cos f)^2.
\end{equation}

The eccentricity of the trajectory obeys
\begin{equation}\label{ecm}
e_{\mathrm{{\mathrm{cm}}}}=1+\frac{2 r_{\mathrm{p}} E_{\mathrm{{\mathrm{cm}}}}}{G M m}
=\sqrt{1+\frac{2E_{\mathrm{cm}}L_{\mathrm{cm}}^{2}}{G^2M^2m^3}},\end{equation}
with $E_{{\mathrm{cm}}}$ the CM-energy,
\begin{equation}
E_{\mathrm{cm}}=\frac{m}{2}\left|\dot{\mathbf{r}}_{\mathrm{{\mathrm{cm}}}}\right|^2-\frac{G M m}{r_{\mathrm{{\mathrm{cm}}}}},
\end{equation}
$a_{\mathrm{{\mathrm{cm}}}}$ the corresponding semi-major axis
\begin{equation}\label{acm}
a_{\mathrm{{\mathrm{cm}}}}=-\frac{GmM}{2 E_{\mathrm{{\mathrm{cm}}}}},
\end{equation}
and $L_{{\mathrm{cm}}}$ the angular momentum
\begin{equation}
\mathbf{L_{{\mathrm{cm}}}}=m \mathbf{r}_{\mathrm{{\mathrm{cm}}}}\times  \mathbf{v}_{\mathrm{{\mathrm{cm}}}},
\end{equation}
perpendicular to the plane of the CM trajectory (aligned to the perpendicular versor $\hat{\mathbf{K}}$ in the $ [\hat{\mathbf{I}},\hat{\mathbf{J}},\hat{\mathbf{K}}]$ centred on the MBH). Note that these expressions (using the appropriate mass) apply equally to the individual components of a binary post-disruption. Fig.~\ref{fig:cartoon_angles} gives a visual representation of our coordinate system.

\subsection{The orbit of the binary}
\label{sec:binary_separation}
\begin{figure}
    \centering
    \includegraphics[width=\columnwidth]{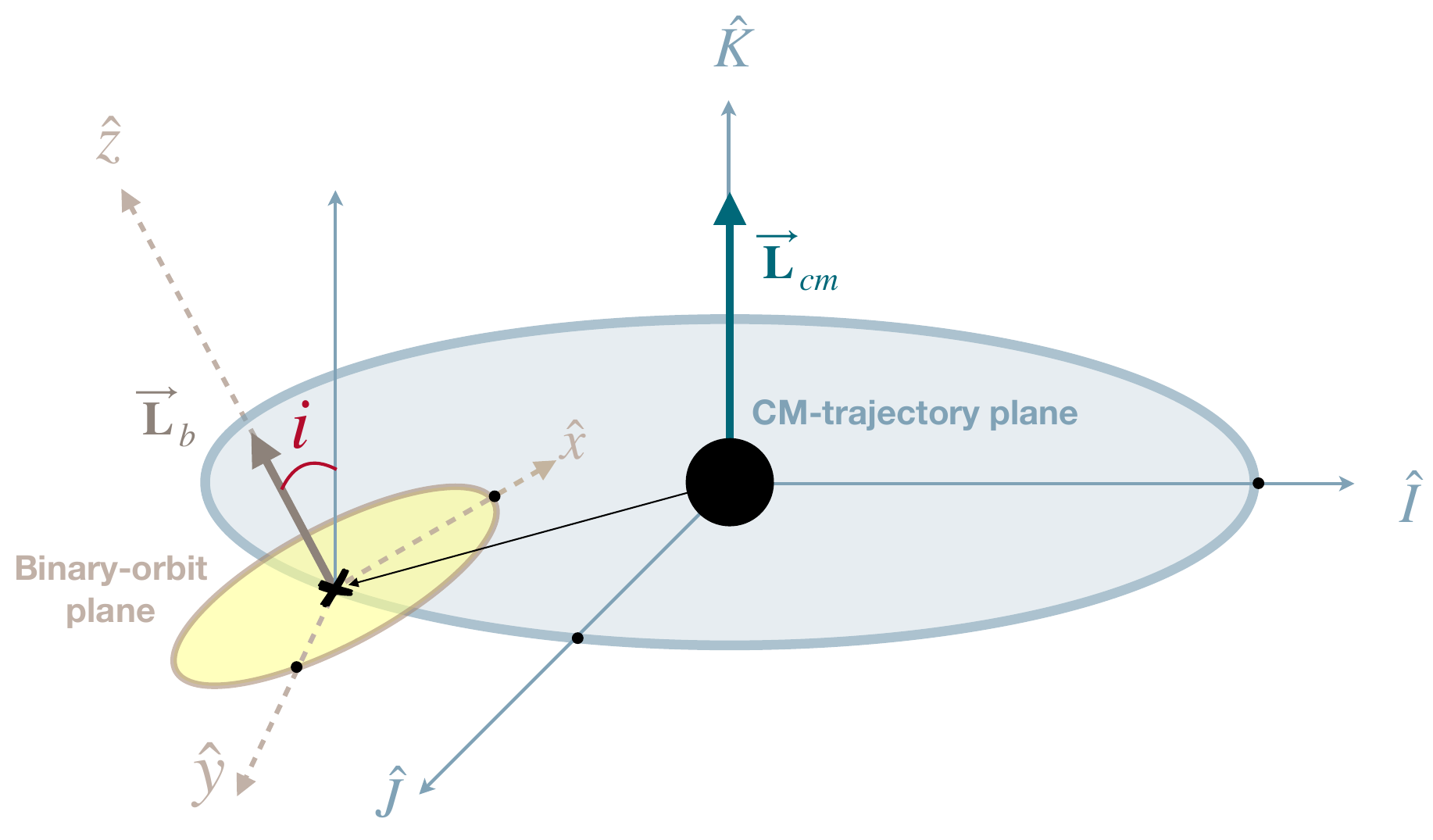}
    \caption{Diagram illustrating the frames of reference used in our calcualtions. $\left[\hat{\mathbf{I}},\hat{\mathbf{J}},\hat{\mathbf{K}}\right]$ (grey arrows) is a coordinate system centred on the MBH, with $\hat{\mathbf{I}}$ and $\hat{\mathbf{J}}$ lying in the CM-trajectory plane (grey ellipse), while $\hat{\mathbf{K}}$ is a versor perpendicular to it and parallel to $\mathbf{L}_{\rm cm}$. A second coordinate system ($\hat{\mathbf{x}},\hat{\mathbf{y}},\hat{\mathbf{z}}$) (ochre dotted arrows) is centred on the binary's CM (at a distance $\mathbf{r}_{\rm cm}$ from the MBH, marked with a black arrow). $\hat{\mathbf{x}}$ and $\hat{\mathbf{y}}$ lie in the binary orbital plane (with $\hat{\mathbf{x}}$ pointing towards the periapsis of the binary orbit - marked with a small black dot). $\hat{\mathbf{z}}$ is perpendicular to it and parallel to the binary total angular momentum $\mathbf{L}_{\rm b}$ (dark-ochre arrow). 
    The inclination $i_0$ is defined as the angle between $\mathbf{L}_{\rm b}$ and its projection along the $\hat{\mathbf{K}}$ direction. 
    }
    \label{fig:cartoon_angles}
\end{figure}

A binary is fully described by its 6 orbital elements: semi-major axis $a_{\rm b}$, eccentricity $e_{\rm b}$, binary phase $\phi$, inclination $i$, argument of periapsis $\omega$ and longitude of ascending node $\Omega$. The last three angles together define the orientation of the binary's orbital plane with respect the CM-trajectory plane. It is useful to work in the frame of the binary orbital plane, defined by unit vectors $\hat{\mathbf{x}}$, $\hat{\mathbf{y}}$ and $\hat{\mathbf{z}}$, where $\hat{\mathbf{x}}$ is in the direction of the pericentre, and $\hat{\mathbf{z}}$ is perpendicular to the plane of the orbit and parallel to the binary angular momentum (see Fig. \ref{fig:cartoon_angles}).
In such frame, the binary motion is described by
\begin{equation}
\mathbf{r}_{\rm b} = \mathbf{r}_2-\mathbf{r}_1 = \frac{a_{\rm b} (1-e_{\rm b}^2)}{1+e_{\rm b} \cos(\phi)}\Big(\cos(\phi)\hat{\mathbf{x}} + \sin(\phi)\hat{\mathbf{y}}\Big)
\end{equation}
and
\begin{equation}
\mathbf{v}_{\rm b} = \mathbf{v}_2-\mathbf{v}_1 = \sqrt{\frac{Gm}{a_{\rm b}(1-e_{\rm b}^2)}} \Big(-\sin(\phi)\hat{\mathbf{x}} + (\cos(\phi)+e_{\rm b})\hat{\mathbf{y}}\Big).
\end{equation}
In the CM-trajectory plane, the following relations hold
\begin{align}
    \begin{split}
\hat{\mathbf{x}}=&\begin{bmatrix}
\cos \Omega \cos \omega - \sin \Omega \cos i \sin \omega \\
\sin \Omega \cos \omega + \cos \Omega \cos i \sin \omega \\
\sin i \sin \omega
\end{bmatrix}, \\
\hat{\mathbf{y}}=&\begin{bmatrix}
-\cos \Omega \sin \omega - \sin \Omega \cos i \cos \omega \\
-\sin \Omega \sin \omega + \cos \Omega \cos i \cos \omega \\
\sin i \cos \omega
\end{bmatrix}, \\
\hat{\mathbf{z}}=&\begin{bmatrix}
\sin \Omega \sin i  \\
- \cos \Omega \sin i \\
\cos i
\end{bmatrix}.
    \end{split}
\end{align}
These can be used to find the initial state of the binary for a given set of initial conditions. Then, as $\mathbf{r}_{\rm b}$ and $\mathbf{v}_{\rm b}$ evolve, we can compute the new orbital elements via
\begin{align}
    E_{\rm b}&=\mu_{\rm b}\frac{1}{2}v_{\rm b}^2-\frac{Gm_1m_2}{r_{\rm b}},\\
    L_{\rm b}&=||\mathbf{L}_{\rm b}|| =\mu_{\rm b} ||\mathbf{r}_{\rm b}\times \mathbf{v}_{\rm b}||,\\
    a_{\rm b}&=-\frac{G m_1 m_2}{2 E_{\rm b}},\\
    e_{\rm b}&=\sqrt{1+\frac{2E_{\rm b} L_{\rm b}^{2}}{G^2 m^2 \mu_{\rm b}^3}},\\
    \cos\phi&=\frac{1}{e_{\rm b}}\left(\frac{a_{\rm b}(1-e_{\rm b}^2)}{r_{\rm b}}-1\right),\\
    \sin\phi&=\frac{1}{e_{\rm b}}\sqrt{\frac{a_{\rm b}(1-e_{\rm b}^2)}{Gm}}\rm (\mathrm{\mathbf{v}}_{\rm b}\cdot \hat{\mathbf{r}}_{\rm b}),
\end{align}
and unit vectors
\begin{align}\label{angles2}
    \begin{split}
     \hat{\mathbf{r}}_{\rm b}&=\mathbf{r}_{\rm b}/||\mathbf{r_{\rm b}}||, \\
       \hat{\mathbf{z}}&=\mathbf{L}_{\rm b}/||\mathbf{L}_{\rm b}||,\\
        \hat{\mathbf{x}}&=\hat{\mathbf{r}}_{\rm b}\cos \phi + (\hat{\mathbf{r}}_{\rm b}\times ,\hat{\mathbf{z}})\sin \phi,\\
        \hat{\mathbf{y}}&=\hat{\mathbf{z}}\times\hat{\mathbf{x}},
    \end{split}
\end{align}
which can be translated into
\begin{equation}\label{angles}
    \begin{aligned}
     i&=\arctan\left(\sqrt{(\hat{\mathbf{z}}[0]^2 + \hat{\mathbf{z}}[1]^2)},\hat{\mathbf{z}}[2]\right)\\
    \omega&=\arctan\left(\hat{\mathbf{x}}[2],\hat{\mathbf{y}}[2]\right)\\
    \Omega&=\arctan\left(\hat{\mathbf{z}}[0],-\hat{\mathbf{z}}[1]\right).
    \end{aligned}
\end{equation}

The inclination is of particular importance as it encodes the direction of the angular momentum of the binary relative to that of the CM trajectory, via $\cos(i)=\mathbf{\hat{L}}_{\rm b}\cdot \mathbf{\hat{L}}_{\rm cm}$ and thus, if the system is prograde ($\cos(i)\rightarrow 1$), retrograde ($\cos(i)\rightarrow -1$) or intermediate ($\cos(i)\sim 0$).

\subsubsection{EOM in the large mass ratio regime}

The large mass ratio ($M / m \gg 1$) ensures the validity of our approximation of a fixed CM trajectory and stationary MBH. We can also simplify the Equation of Motion (EOM) governing the relative motion of the two binary members, $\mathbf{r}_{\rm b}$. Following \citet[][]{Kobayashi12}, we start by considering the EOM of each binary member separately,
\begin{equation}
\begin{aligned}
& \ddot{\mathbf{r}}_1=-\frac{G M}{r_1^3} \mathbf{r}_1+\frac{G m_2}{\left|\mathbf{r}_1-\mathbf{r}_2\right|^3}\left(\mathbf{r}_2-\mathbf{r}_1\right), \\
& \ddot{\mathbf{r}}_2=-\frac{G M}{r_2^3} \mathbf{r}_2-\frac{G m_1}{\left|\mathbf{r}_1-\mathbf{r}_2\right|^3}\left(\mathbf{r}_2-\mathbf{r}_1\right).
\end{aligned}
\end{equation}
Then, the equation for the distance between the two stars is
\begin{equation}\label{Eom_rel}
\ddot{\mathbf{r}}_{\rm b}=-\frac{G M}{r_2^3} \mathbf{r}_2+\frac{G M}{r_1^3} \mathbf{r}_1-\frac{G m}{r_{\rm b}^3} \mathbf{r}_{\rm b}.
\end{equation}
We now assume that the distance between the two stars is much smaller than that to the MBH and thus linearise the first two terms of Eq. (\ref{Eom_rel}) around the position of the binary CM ($\mathbf{r}_{\mathrm{cm}}$), 
\begin{equation}\label{linearised}
    \ddot{\mathbf{r}}_{\rm b}=-\frac{G M}{r_{\mathrm{cm}}^3} \mathbf{r}_{\rm b}+3 \frac{G M}{r_{\mathrm{cm}}^5}\left(\mathbf{r}_{\rm b} \mathrm{r}_{\mathrm{cm}}\right) \mathbf{r}_{\mathrm{cm}}-\frac{G m}{r_{\rm b}^3} \mathbf{r}_{\rm b}+\mathcal{O}\left(\left(\frac{\mathbf{r}_{\rm b}}{\mathbf{r_{\mathrm{cm}}}}\right)^2\right).
\end{equation}
We can rescale this equation in terms of a characteristic length scale $\lambda=(m / M)^{1 / 3} r_{\rm p}$ and time scale $\tau=\sqrt{r_{\rm p}^3 / G M}$. The linearized EOM can be rewritten in terms of the dimensionless variable $\boldsymbol{\eta} \equiv\frac{1}{\lambda}\mathbf{r}_{\rm b}$ and the shorthand for the derivative $g'= \tau\dot{g}$ giving
\begin{equation}\label{EOM}
\boldsymbol{\eta}''=\left(\frac{r_{\rm p}}{r_{\mathrm{{\mathrm{cm}}}}}\right)^3\left[-\boldsymbol{\eta}+3\left(\boldsymbol{\eta} \cdot \hat{\mathbf{r}}_{\mathrm{{\mathrm{cm}}}}\right) \hat{\mathbf{r}}_{\mathrm{{\mathrm{cm}}}}\right]-\frac{\boldsymbol{\eta}}{|\boldsymbol{\eta}|^3}+\large{\mathcal{O}}\left(\left(\frac{\boldsymbol{\eta}}{\boldsymbol{r}_{\mathrm{cm}}}\right)^2\right).
\end{equation}
Setting $\boldsymbol{\eta}=(\eta_x, \eta_y, \eta_z)$, we explicitly rewrite equation (\ref{EOM}) at first order in dimensionless Cartesian coordinates,
\begin{equation}\label{EOMs}
\begin{aligned}
\eta_x'' \approx & \frac{(1+e_{\mathrm{{\mathrm{cm}}}} \cos f)^3}{(1+e_{\mathrm{{\mathrm{cm}}}})^3}[-\eta_x+3(\eta_x \cos f+\eta_y \sin f) \cos f] \\
& -\frac{\eta_x}{\left(\eta_x^2+\eta_y^2+\eta_z^2\right)^{3 / 2}}, \\
\eta_y'' \approx & \frac{(1+e_{\mathrm{{\mathrm{cm}}}} \cos f)^3}{(1+e_{\mathrm{{\mathrm{cm}}}})^3}[-\eta_y+3(\eta_x \cos f+\eta_y \sin f) \sin f] \\
& -\frac{\eta_y}{\left(\eta_x^2+\eta_y^2+\eta_z^2\right)^{3 / 2}}, \\
\eta_z'' \approx & -\frac{(1+e_{\mathrm{{\mathrm{cm}}}} \cos f)^3}{(1+e_{\mathrm{{\mathrm{cm}}}})^3} \eta_z-\frac{\eta_z}{\left(\eta_x^2+\eta_y^2+\eta_z^2\right)^{3 / 2}}.
\end{aligned}
\end{equation}
Numerical integrations can be performed in these general coordinates, and the physical variables can be recovered by reintroducing the dimensionally consistent combination of characteristics scales (e.g. $\mathbf{v}_{\rm b} = \frac{\lambda}{\tau} \boldsymbol{\eta}'$). 

The relationship between time, $t$, and the CM phase, $f$, depends on the trajectory under consideration; the results for a bound, parabolic and hyperbolic trajectory are, respectively,  
\begin{equation}
t/\tau=
\begin{cases}
(1-e_{\mathrm{{\mathrm{cm}}}})^{-3 / 2}(\xi-e_{\mathrm{{\mathrm{cm}}}} \sin \xi)  \quad &\text{for}\quad E_{\mathrm{cm}}<0\\
\sqrt{2}\left(\xi+\xi^3 / 3\right)  \quad &\text{for}\quad E_{\mathrm{cm}}=0\\
 (e_{\mathrm{{\mathrm{cm}}}}-1)^{-3 / 2}(e_{\mathrm{{\mathrm{cm}}}} \sinh \xi-\xi) \quad &\text{for}\quad E_{\mathrm{cm}}>0,
\end{cases}
\end{equation}
where the eccentric anomaly $\xi$ is related to the true anomaly according to
\begin{equation}
    \xi=\begin{cases}
        \arctan\left(\frac{\sqrt{1-e_{\mathrm{{\mathrm{cm}}}}^2}\sin f}{e_{\mathrm{{\mathrm{cm}}}}+\cos f}\right)\quad &\text{for}\quad E_{\mathrm{cm}}<0\\
        \tan(f/2) \quad &\text{for}\quad E_{\mathrm{cm}}=0\\
        \ln \left(\frac{\sqrt{e_{\mathrm{{\mathrm{cm}}}}+1}+\sqrt{e_{\mathrm{{\mathrm{cm}}}}-1} \tan(f/2)}{\sqrt{e_{\mathrm{{\mathrm{cm}}}}+1}-\sqrt{e_{\mathrm{{\mathrm{cm}}}}-1} \tan (f/2)}\right)\quad &\text{for}\quad E_{\mathrm{cm}}>0.\\
        
    \end{cases}
\end{equation}
\citep[see e.g.][]{Landau,Murray} where $t=0$ at periapse.

\subsubsection{Diving factor}
\label{sec:units}
One of the main parameter of our analysis is the dimensionless diving factor, defined as
\begin{equation}
   \beta=\frac{r_{\rm t}}{r_{\rm p}},
\end{equation}
which quantifies how deeply into the tidal sphere the binary can dive: i.e. $\beta<1$ corresponds to shallow encounters outside the tidal sphere of influence, while  $\beta>1$  corresponds to deeper encounters  within it. The tidal radius, $r_{\rm t}$, depends on $a_{\rm b}$, which varies throughout the interaction, thus we will use the initial value $\beta_0$ to parameterize an interaction, defined it in terms of the initial value of $a_{\rm b,0}$. Specifying the diving factor allows us to remove the degenerate term $\frac{r_{\rm p}}{a_{\mathrm{b},0}}$ from the intial conditions.

As we saw earlier in this section, it is convenient to rescale lengths and times in terms of $\lambda$ and $\tau$, which can be re-expressed in terms of the properties of the binary and the diving factor as
\begin{equation}\label{lambda}
\lambda=Q^{-1 / 3} r_{\rm p}=\beta_0^{-1}a_{\mathrm{b},0},
\end{equation}
and 
\begin{equation}\label{taulambda}
\tau=\sqrt{\frac{r_{\rm p}^3}{ G M}} =  \sqrt{\frac{\lambda^3}{Gm}} = \beta_0^{-\frac{3}{2}}\frac{P_{\mathrm{b},0}}{2\pi}
\end{equation}
where $P_{\rm b}=2\pi \sqrt{\frac{a_{\rm b}^3}{Gm}}$ is the period of the binary. Hence for $\beta\sim 1$ the characteristic scales of the problem are approximately the characteristic length, time and mass scales of the binary.

In summary, the full behavior of the system can be captured by the initial parameters $Q, q, \beta_0$, $e_{\rm cm}$, $e_{\rm b,0}$, $\phi_0$, $i_0$, $\omega_0$, $\Omega_0$ and $t_0$ (and then scaled to physical units by specifying, for example, $m$ and $a_{\rm b,0}$).

\section{The first pericentre passage}\label{distr}
In this section, we illustrate the numerical integration procedure followed to analyse a single encounter between a binary and the MBH.

\begin{figure}
    \centering
\includegraphics[width=\columnwidth]{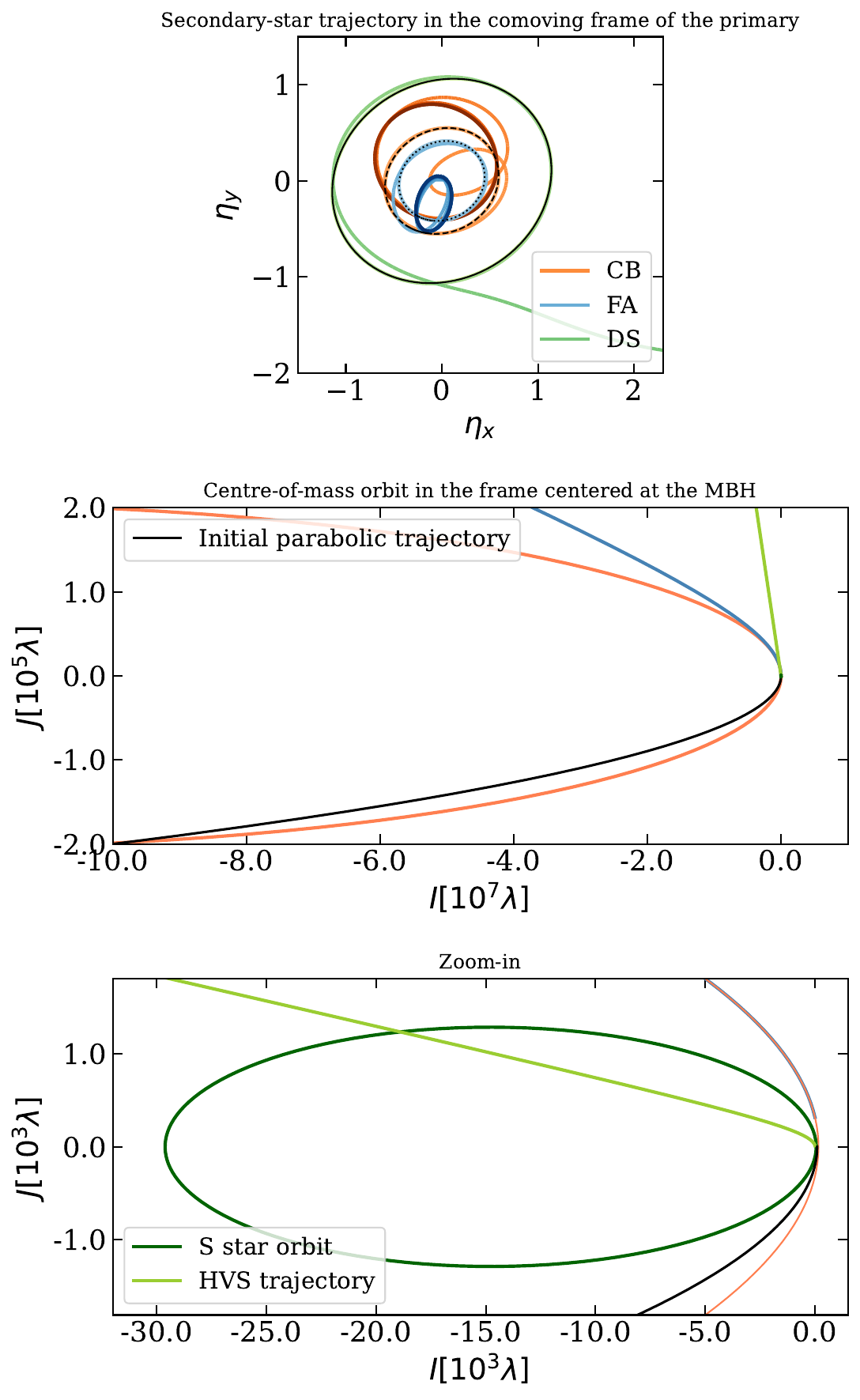}
   \caption{Examples of orbits for a CB (orange), a FA (blue) and a disrupted binary (green) obtained from initially-circular binaries on parabolic trajectories for a set of phases and angular parameters sampled as described in the text.  In all panels coordinates are expressed in code units (see \ref{sec:units}) 
   \emph {Upper panel}:  secondaries' orbits in the comoving frame of their respective primaries (colors change from lighter to darker as the system evolves). For clarity, the initial trajectories are marked in black (dashed line for CBs, dotted line for FAs and solid line for Ds).
   \emph{Centre panel:} trajectories of the CMs  after tidally interacting with the MBH. In black the initial parabolic CM-trajectory common to all the binaries.  \emph{Lower panel}: Zoom in on the captured and ejected stars' trajectories.}
    \label{fig:traj}
\end{figure}

\begin{figure*}
\includegraphics[width=\textwidth]{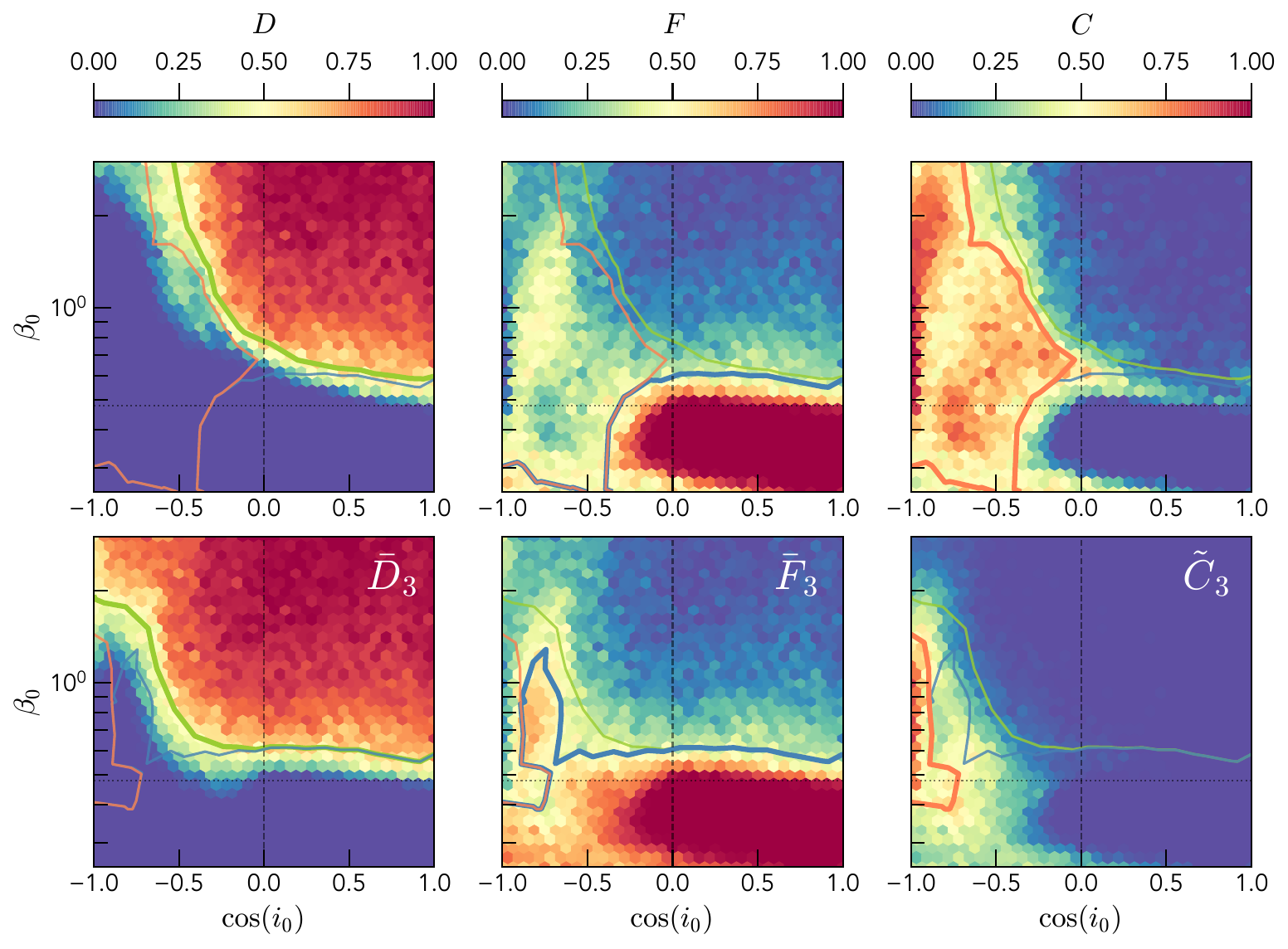}
    \caption{Fractions of disruptions (D), fly-aways (F) and coming-backs (C) as a function of the diving factor $\beta_0$ and the initial inclination $i_0$. The green, blue and orange contour lines highlight the regions of parameter space where D,F and C, are, respectively, at least 0.5.
    Lines are thicker when the contour is in the corresponding panel. This figure is generated for 100,000 interactions with random initial conditions, except for $e_{\mathrm{b},0}=0$ (initially circular binary) and $e_{\rm cm}=1$ (on a parabolic orbit).
    The vertical dashed line corresponds to an inclination of $\frac{\pi}{2}$.
    The dotted horizontal line is at $\beta_{\text{lim}}=0.4779703$, the smallest value at which an initially circular binary can disrupt.
    \emph{Top row}: Fractions after a single passage.  The equivalent behavior for non-circular initial binaries can be seen in Fig. \ref{hexbinHCF_ecc}. 
    \emph{Bottom row}: Overall fractions after 3 passages (as detailed in Section \ref{Multiplep}). The remnant fraction of comebacks ($\tilde{C}_3$) will eventually add to either Ds or FAs.
    }
    \label{fig:hexbinHCF}
\end{figure*}

\subsection{Initial conditions and orbit integration}\label{orbitint}
We need to choose an initial time early enough that the tidal influence of the MBH is initially minor, but late enough that the simulation only has to run for a few binary periods until the tidal influence starts to become significant. Hence we choose $t_0$ ($<0$ as periapse occurs at $t=0$) such that
\begin{equation}
t_0= \begin{cases}-(t_{r_\mathrm{t}}+N P_{\rm b}) & \text { if } \beta_0 \geq 1 \;\mathrm{and} \;t_{r_\mathrm{t}}>t_{\frac{\pi}{2}},\\ -(t_{\frac{\pi}{2}}+N P_{\rm b}) & \text {else} \end{cases}
\end{equation}
where $t_{r_{\rm t}}$ is the (positive) time at which the CM trajectory passes through the tidal radius, $t_{\frac{\pi}{2}}$ is the time at which $f=\frac{\pi}{2}$, and $N$ is approximately the number of binary periods before each of these times occur. For systems which do not pass through the tidal radius ($\beta_0 \leq 1$) or do so just before periapse ($t_{r_{\rm t}} < t_\frac{\pi}{2}$) starting the simulation slightly before $f=-\frac{\pi}{2}$ captures much of the curvature of the CM trajectory without overly-long integration times. Generally we have found $N=3$ to be sufficient to capture the full interaction. This recipe for choosing a suitable $t_0$ (as a function of $\beta_0$ and $e_{\rm cm}$) essentially removes another initial parameter for our integration.

The initial position and velocity in our rescaled coordinates, $\boldsymbol{\eta}_0$ and $\boldsymbol{\eta}_0'$ can be found from the initial orbital elements using the relationships defined in section \ref{sec:binary_separation}, converting using $a_{\mathrm{b},0} = \beta_0 \lambda$ and $\sqrt{\frac{Gm}{a_{\mathrm{b},0}}}=\beta_0^{-1} \frac{\lambda}{\tau}$.

We integrate the system of Eqs. (\ref{EOMs}) using the scipy function \texttt{odeint}. Due to the presence of typically very different characteristic time-scales we choose timesteps $dt =\varepsilon \min(P_{\rm b,0},t_{\rm dyn})$ where 
\begin{align}\label{period}
    t_{\text{dyn}}(f)&= 2\pi\sqrt{\frac{r_{\mathrm{{\mathrm{cm}}}}^{3}}{G M}}=2\pi\left(\frac{1+e_{\mathrm{{\mathrm{cm}}}}}{1+e_{\mathrm{{\mathrm{cm}}}} \cos(f)}\right)^{3/2}\tau,
\end{align}
is the dynamical time for a given $f$.
$\varepsilon$ is a small factor chosen to balance sufficient accuracy and low computational cost, for which we find $\varepsilon=0.01$ to be a generally suitable choice. In general the binary period is small far from periapse (and thus sets the timestep there) but the dynamical time defines the behavior near periapse. 

Comparison simulations performed with a full adaptive n-body integrator, \texttt{REBOUND} \citep[][]{ReinLiu2012,ReinSpiegel2015}, shows in general excellent agreement, validating both the numerical method and our analytically simplified EOM.

\subsection{Outcomes of the interaction}


As the CM trajectory is taken to be fixed, the associated energy, $E_{\rm cm}$, is constant. Over the integration the internal energy of the binary changes by
\begin{equation}\label{enexchangeb}
\Delta E_{\rm b}=E_{\mathrm{b},f}-E_{\mathrm{b},i}.
\end{equation}
From this we can approximate the transfer of tidal energy between the CM trajectory and the binary orbit by conservation of energy, and thus
\begin{equation}\label{enexchange}
E_{\mathrm{cm},f}=E_{\rm cm}-\Delta E_{\rm b},
\end{equation}
with $E_{\mathrm{b},i}<0$ (and $E_{\rm cm}$ taken to be 0 in the case of a parabolic trajectory).
If the binary final energy is positive (that is $E_{\mathrm{b},f}>0$), then the binary will be unbound; we call these systems \textit{disruptions} (Ds). 
If instead the binary final energy is still negative, then the binary will remain bound. Now there can be two outcomes: if $E_{\mathrm{cm},f}>0$ the the CM is on an unbound trajectory, and these we call \textit{fly-away} binaries (FAs); if $E_{\mathrm{cm},f}<0$, the CM is bound to the MBH and the system will return for a subsequent passage. We call these \textit{come-back} binaries (CBs). 

In Fig. \ref{fig:traj} we provide some examples of interactions representing the three possible fates for a binary: a dirsupted system with $\beta_0=1.17$, a fly-away binary with $\beta_0=0.46$, and a come-back binary with $\beta_0=0.61$. Otherwise the initial conditions are all the same with initially circular binaries ($e_{\mathrm{b},0}=0$) on parabolic CM-trajectories ($e_{\mathrm{cm}}=1$), initial orientation $i_0=\omega_0=\Omega_0=\pi/2$, and binary phase $\phi_0=\pi$.

For parabolic CM trajectories ($E_{\rm cm}=0$) the eventual fate depends entirely on whether the binary gains or loses energy: $E_{\mathrm{cm},f}=-\Delta E_{\rm b}$. Thus, surviving binaries that shrink ($a_{\mathrm{b},f}<a_{\mathrm{b},i}$) will be FAs while if they become larger ($a_{\mathrm{b},f}>a_{\mathrm{b},i}$) they will be CBs. 

Figure \ref{fig:hexbinHCF} shows the likelihood of a given outcome as a function of $\beta_0$ and $i_0$ for initially circular binaries on parabolic orbits (marginalized over $\phi_0, \Omega_0$ and $\omega_0$ chosen uniformly and randomly between $0$ and $2\pi$).

According to previous literature \citep[see again][]{Sari10,Brown18,Kobayashi12}, the highest (lowest) fraction of ejected stars is produced by initially prograde (retrograde) binaries with at least $\beta_\text{lim}\equiv0.4779703$, while for smaller $\beta_0$ there are no disruptions of initially circular binaries.

Now with a full range of inclinations we can further analyze the possible outcomes. We can generally divide the behavior between a strongly \textit{prograde} regime ($1\leq \cos(i_0) \lesssim \frac{1}{3}$), and intermediate regime ($\frac{1}{3} \lesssim \cos(i_0) \lesssim -\frac{1}{3}$) and a strongly retrograde regime ($-\frac{1}{3} \lesssim \cos(i_0) \leq -1$).
We can see that Ds are prevalent in the prograde regime for $\beta_0>\beta_{\rm lim}$ but that this behavior extends to intermediate inclinations at slightly higher $\beta_0$, including frequent disruptions for weakly prograde systems ($0 < \cos(i_0) \lesssim -\frac{1}{3}$) for $\beta_0 \gtrsim 1$. Strongly retrograde systems are much more resistant to disruption, with the fraction never reaching 50$\%$.

One way to understand the dependence on inclination is to think about the timespan over which one member of the binary is the closest to the MBH. Prograde binaries rotate with the trajectory, and thus this timespan is longer, and the inverse is true for retrograde systems. Thus, for more prograde binaries, the tidal force is acting on the binary in a consistent sense for longer. Thus a larger tidal force (deeper encounter) is needed to have the same time-integrated effect and disrupt more retrograde systems.

Below $\beta_{\rm lim}$ almost all prograde and intermediate systems are FAs; these also occur occasionally for retrograde orbits and infrequently across the remaining part of the parameter space. Initially retrograde systems are most likely to end up as CAs, even at large $\beta_0$s. Our simulations only cover $\beta_0>0.25$ but there is putative evidence that $F$ and $C$ tend to 0.5 (i.e. both outcomes equally likely) for shallower interactions, independent of the inclination.

In summary, the three outcomes broadly occupy different parts of the parameter space: 
\begin{itemize}
    \item Ds tend to occupy the area where $\cos(i_0)\gtrsim -1/3$ and $\beta_0 > \beta_{\rm lim}$,
    \item FAs the area where $\cos(i_0)\gtrsim -1/3$ and $\beta_0\lesssim 0.6$,
    \item CBs spans the entire $\beta_0$ range preferentially for $\cos(i_0)\lesssim -1/3$. 
\end{itemize}

We also perform the same analysis on non-circular initial binaries in appendix \ref{app:ecc}, with broadly the same conclusions - excepting that eccentric binaries can disrupt with $\beta_0<\beta_{\rm lim}$ and that the overall fraction of FAs is reduced.

\begin{figure}
    \centering
    \includegraphics[width=\columnwidth]{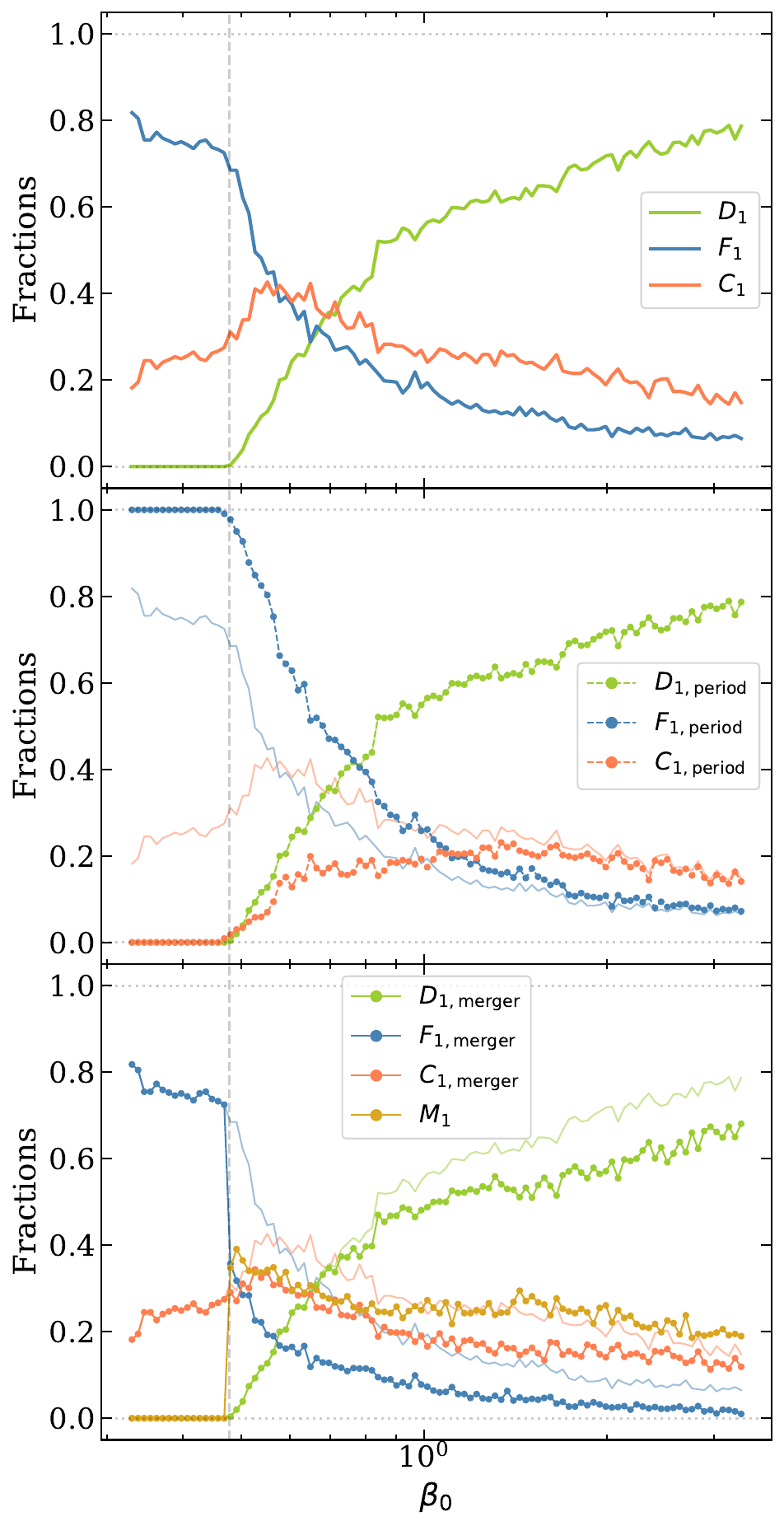}
    \caption{\emph{Upper panel}: Fractions of Ds, FAs and CBs (green, blue and orange, respectively), after one pericentre passage.
    \emph{Central panel}:
    Comparison between the fractions of Ds, FAs and CBs before (thin lines, same as in the upper panel) and after period cut (dotted lines, same colors) as detailed in section \ref{sec:lifetime1} assuming an example binary with $a_{\mathrm{b},0}=0.1$ AU, $m=4 M_{\odot}$ and $q=\frac{1}{3}$. 
    \emph{Lower panel}:
    Comparison between the fractions of Ds, FAs and CBs before and after accounting for mergers (dotted lines, same colors) as detailed in section \ref{sec:mergers1} assuming the same example binary.
    The gold dotted lines marks the fraction of Ms after one passage.}
    \label{fig:HCFbothcuts}
\end{figure}

In Fig. \ref{fig:HCFbothcuts} (upper panel) we show the fraction of each outcome as a function of $\beta_0$, marginalized over $\cos(i)$, i.e. the average for random binary orientations. As expected, Ds dominate at high $\beta_0$s (and are still rising for our maximum value of 3.3), though, as shown in \citet{Sari10}, even at very high $\beta$ some small fraction of binaries can survive. FAs are most common for $\beta\lesssim 0.6$ ($\sim 80 \%$ of outcomes) and drop off steeply for deeper encounters. CBs are almost always sub-dominant but consistently account for $\gtrsim 20 \%$ of the outcomes.

\begin{figure}
    \centering
\includegraphics[width=\columnwidth]{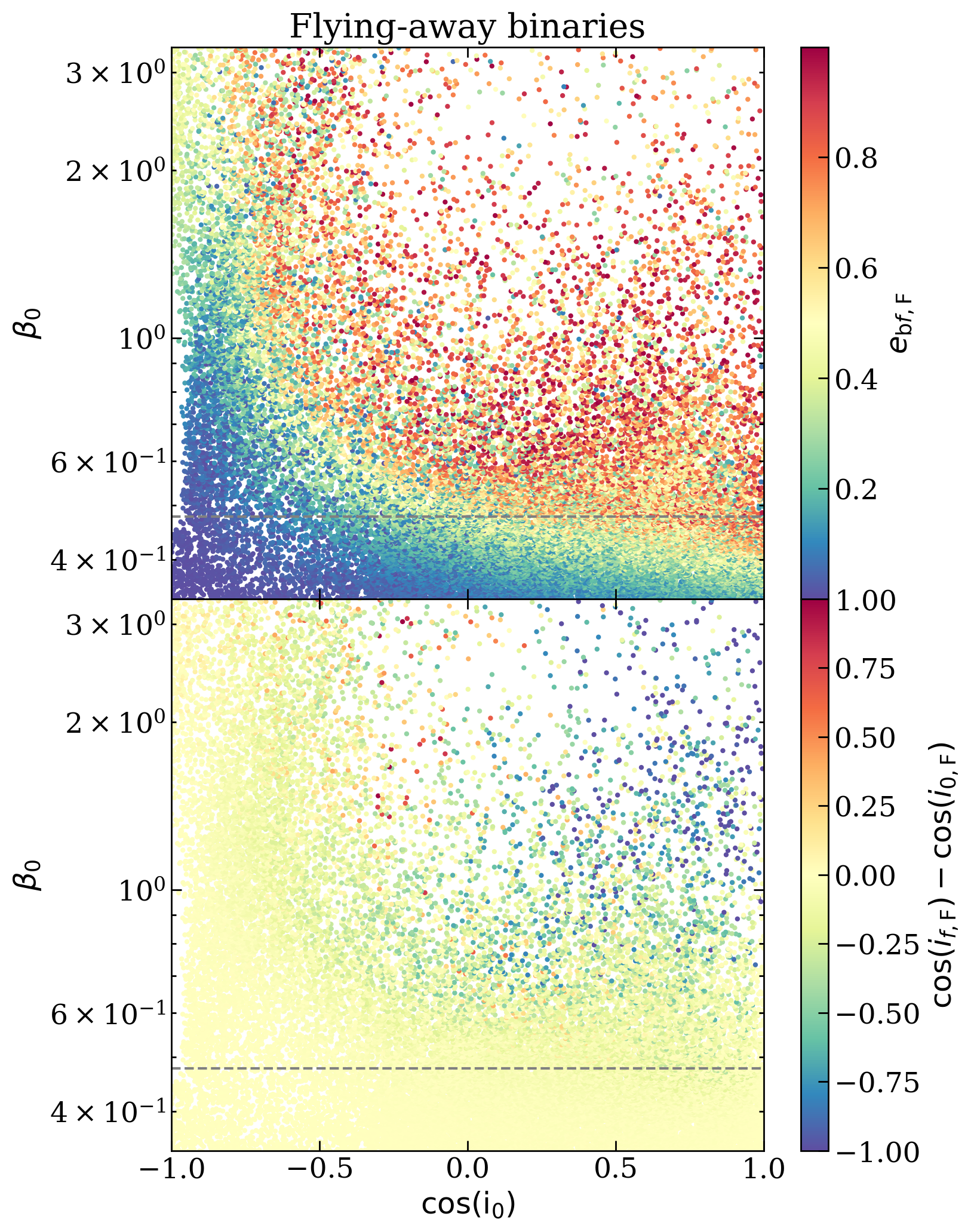}
    \caption{Set of FAs resulting from one interaction between an initial population of $100000$ circular binaries on a parabolic orbit and the MBH in the $\beta_0$-$\cos(i_0)$ plane (marginalized over $\omega$, $\Omega$ and the binary phase), coloured by the final FA binary eccentricity (top panel) and by the difference between their final ($\rm i_{f,\rm F}$) and initial ($\rm i_{0,\rm F}$) binary inclinations (bottom panel).}
    \label{fig:propF1}
\end{figure}

\begin{figure}
    \centering
    \includegraphics[width=\columnwidth]{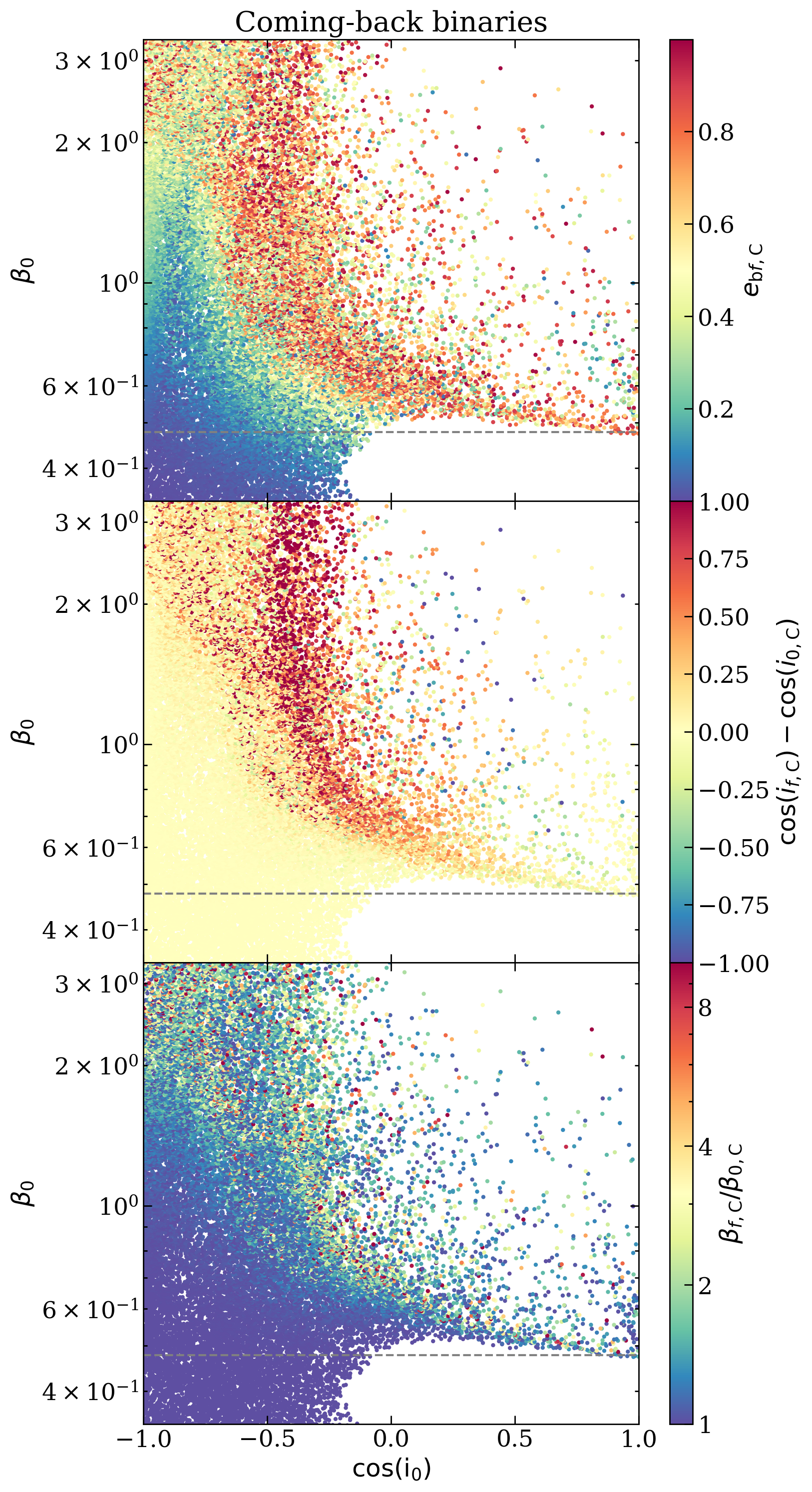}
    \caption{Similar to Fig. \ref{fig:propF1}, we show the properties of CBs coloured by the final binary eccentricity (top panel), the difference between their final ($\rm i_{f,\rm C}$) and initial binary inclinations ($\rm i_{0,\rm C}$) (centre panel) and the ratio between their final ($\rm \beta_{f,\rm C}$) and initial ($\rm \beta_{0,\rm C}$) diving factors (bottom panel).}
    \label{fig:propC1}
\end{figure}

For a more complete characterization of the properties of FAs and CBs after the encounter, we now refer to Figures (\ref{fig:propF1},\ref{fig:propC1}), respectively.
Many surviving binaries have significant eccentricity, with only strongly retrograde and low $\beta_0$ systems with $e_{\mathrm{b},f}<0.2$. As regards the change in inclination, we see a clear difference between FAs, which tend to become more retrograde, and CBs, which mostly become more prograde. 
Finally for CBs we can ask what the initial $\beta$ of their next interaction will be. In general, the CM trajectory is only marginally altered (e.g. the CM eccentricity changes by less than $1\%$),
and thus the dominant change is the enlarged $a_{\rm b}$, causing essentially all CBs to come back with larger $\beta$ (though only marginally so in many cases).
Combining these observations we can see that CBs tend to come back with higher $\beta$s, more prograde and more eccentric. Thus, they generally move up and right in $\beta, \cos(i)$ space towards the region where Ds are more likely. Even where the change is marginal, they are moving towards the region where the next encounter will cause greater changes in their properties and thus, uninterrupted, may be expected to evolve eventually towards disruption.

\subsection{Physical constraints}
\label{sec:phys_constraints}

The results presented so far are general and, therefore, applicable to any kind of binary, independent of any physical properties of the system, encoded in the underlying rescalings and characteristic units. However, the channels presented so far can be influenced by the star properties; in particular, we consider their finite lifetime and size. The former can lead to CBs with a long CM trajectory period to not survive until the next encounter. The latter can lead to mergers. 
Smaller separation systems will have a smaller characteristic dynamical timescales compared to their lifetime, and smaller radii objects will be less likely to interact tidally - thus our previous conclusions can be expected to hold for tight compact object binaries but may be increasingly affected for longer period stellar objects.

In the next two sub-sections, we analyze these alternative fates, by choosing an example binary (thus setting the physical time and length scales) with $a_{\mathrm{b},0}=0.1$ AU, $m=4 M_\odot$ (and thus for SgrA* $Q=10^6$), $q=\frac{1}{3}$ and stellar radii which obey $R/R_\odot \sim M/M_\odot$.

\subsubsection{Impact of stellar lifetime on coming-back binaries}
\label{sec:lifetime1}

If the lifetime of either star is shorter than the CM trajectory period, these binaries can be considered to be FAs (in that they will not experience a subsequent passage). In most cases we can expect that the more massive primary has the shorter lifetime. 

The main sequence lifetime of the primary in our example system ($m_1=3 M_\odot$) is of order 100 Myr. In theory a system may not undergo a Hills mechanism encounter until part way through their life, giving a more stringent constraint, but we will ignore this for our simple order of magnitude analysis.

We will compare the stellar lifetime with the period of the CM trajectory of CBs around the MBH, which is\footnote{The definition we are using to compute the period is valid for orbits in a Kepler potential. However, if the binary travels outside the MBH sphere of influence, our description is no longer completely valid. Including the excess mass enclosed at larger distances would reduce $P_{\rm cm}$. Thus, in this respect our cut is conservative, in that we use the maximum possible period.}
\begin{equation}
    P_{\rm cm}=2\pi\sqrt{\frac{a_{\mathrm{cm}}^{3}}{GM}}.
\end{equation}

The smallest possible $P_{\rm cm}$ corresponds to a CM trajectory with the largest possible (negative) $E_{\rm cm}$. Assuming an initially  parabolic CM trajectory $E_{\rm cm} = -\Delta E_{\rm b}$ and the largest possible $\Delta E_{\rm b}$ is just less than $E_{\rm b,0}$ (the binary is almost but not quite unbound). In this case
\begin{equation}
a_{\rm cm, min} = -\frac{GMm}{2 E_{\mathrm{b},0}} = \frac{Mm}{m_1 m_2}a_{\rm b,0}  = Q\frac{(1+q)^2}{q}a_{\rm b,0} 
\end{equation}
and thus
\begin{equation}
P_{\rm cm, min} = Q \frac{(1+q)^3}{q^\frac{3}{2}} P_{\rm b,0}.
\end{equation}
There is no maximum period as $\Delta E_{\rm b}$ can be vanishingly small, but we can reasonably expect a characteristic period of CBs to be within a few orders of magnitude of $P_{\rm cm, min}$.

\begin{figure}
    \centering
    \includegraphics[width=\columnwidth]{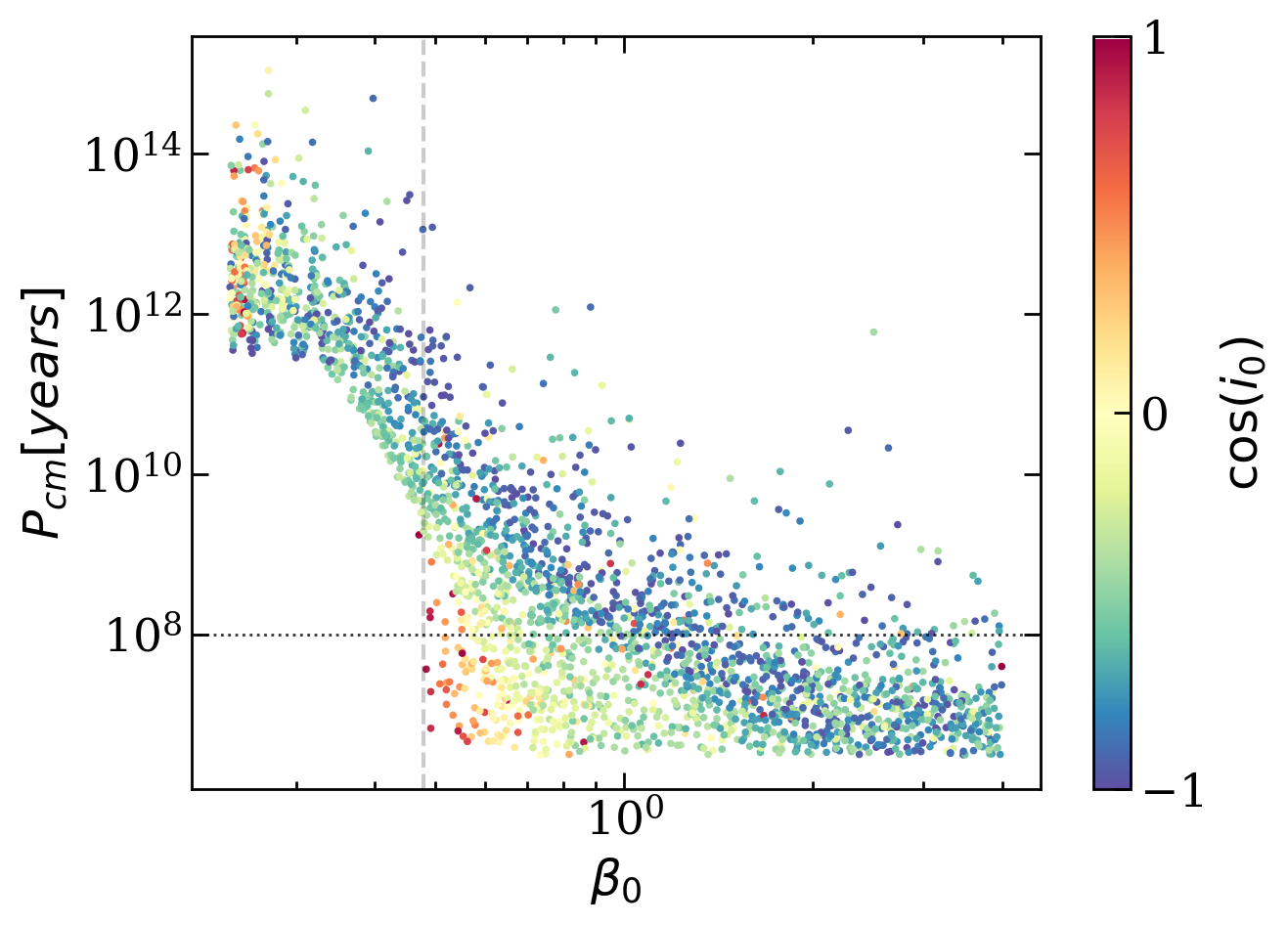}
    \caption{Distributions of the periods of the CM-trajectory of CBs as a function of $\beta_0$, coloured by the cosine of the initial binary inclination ($i_0$). 
    The vertical dashed line corresponds to $\beta_{\rm lim}$. The horizontal dotted black line marks periods of the order $10^{8}$ years (approximate life-time of MS stars).
   }
    \label{periods}
\end{figure}

In Figure \ref{periods}, we show the periods of CBs from an initial population of 100,000 randomly oriented circular binaries. Below $\beta_{\rm lim}$ the energy exchange is minimal ($\ll E_{\rm b,0}$) and the periods are generally very long. Above $\beta_{\rm lim}$ the change in energy is much more significant, especially for more prograde systems, and we see periods approaching $P_{\rm cm, min}$. The median $P_{\rm cm}$ drops below $10^8$ years for $\beta_0 \approx 0.75$.

We can see the total effect in the middle panel of Fig. \ref{fig:HCFbothcuts}. As we would expect the fraction of CBs is significantly reduced, with almost none below $\beta_{\rm lim}$, with the effect reducing at higher $\beta$. The fraction of FAs is raised by the same amount (and Ds are unaffected).

\subsubsection{Impact of mergers}\label{Mergers}
\label{sec:mergers1}

Taking into account the finite size of the binary components allows us to account for the possibility that components of a binary merge. We use \textit{merger} to refer to any time the two binary members come close enough to lose significant energy to tidal deformation. These systems would evolve internally in a way not captured by our integration. In some cases this may result in true collisions, in others possible mass transfer and tight tidally circularized binaries.

Including mergers means introducing a new channel, which will be referred to as {\it Ms}. In this section, we quantify the fraction of mergers after one pericentre passage and assess the consequent impact on the fractions of Ds, CBs and FAs. 

Practically, we consider a binary star as ``merged'' when the primary fills its Roche lobe \citep[see][]{Eggleton1983}, 
\begin{equation}
    R_{\rm L,1}=\frac{0.49q^{2/3}}{0.6q^{2/3}+\ln(1+q^{1/3})}\ r,
\end{equation}
and become tidally deformed, i.e. when $R_{\rm L,1}\leq R_1$.
We note that $R_{\rm L,1}$ amounts to $\approx0.289 \ r$ for $q={1/3}$.

Thus a system is considered to merge if at any point the binary separation $r \lesssim r_{\rm merge}$ where $R_{L,1}(r_{\rm merge})=R_1$.
Using a simple proxy for the radius of a $\sim$stellar mass star of $R/ R_\odot = M/M_{\odot}$ this means our example system has $r_{\rm merge}\approx 0.048$ AU $\approx \frac{1}{2}a_{\mathrm{b},0}$.

\begin{figure}
    \centering
    \includegraphics[width=\columnwidth]{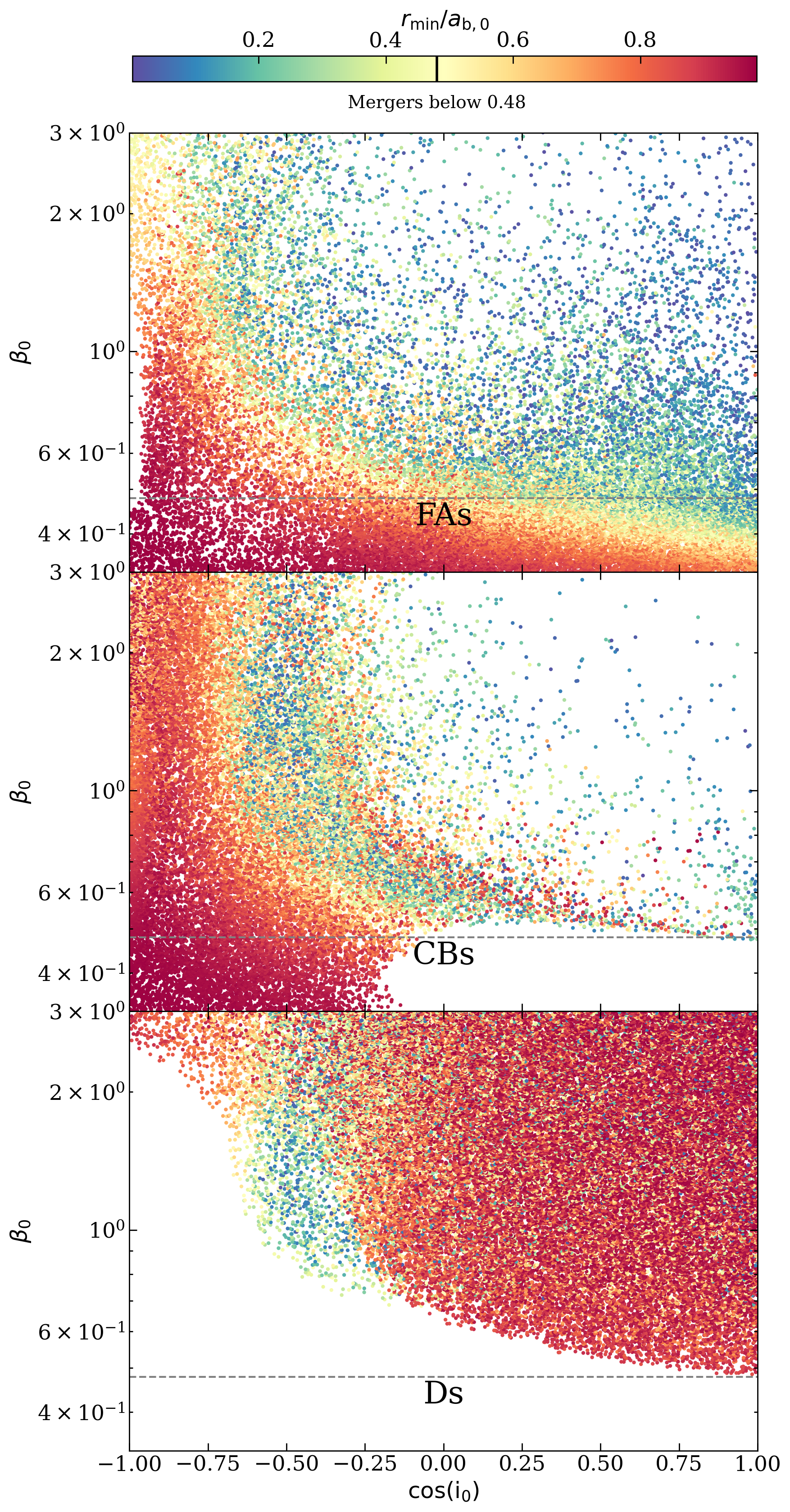}
    \caption{Set of $100000$ initially-circular binaries on a parabolic orbit in the $\beta_0$-$\cos(i_0)$ plane, coloured by the ratio between the minimum dimensionless distance $\rm r_{min}$ and the initial binary semi-major axis $a_{\rm b0}$, after one pericentre passage (marginalized over $\omega$, $\Omega$ and the binary phase) and divided into FAs (top panel), CBs (central panel) and Ds (bottom panel). For our choice of the binary and our definition of the merger condition (see Sec. \ref{sec:mergers1}), mergers happen for $r_{\rm min}/a_{\mathrm{b},0}\leq 0.48$, marked on the colorbar.}
    \label{fig:rmin}
\end{figure}

The minimum radii that our simulated systems reach are shown in Fig. \ref{fig:rmin}. Binaries that contribute to different channels (FAs, CBs, Ds) reach the smallest $\rm r_{\rm min}/a_{\rm b,0}$ in the same part of the $\cos(i_0),\beta_0$ parameter space: the blue stripe which is most evident in the second panel. The exception for this is that for FAs there is no upper limit where $\rm r_{\rm min}/a_{\rm b,0}$ returns to $\sim 1$. The behavior is relatively uniform up to this stripe, and then much more varied above it - which as previously suggested may be due to encounters where the binary undergoes multiple periods of deformation. 

Returning now to the last panel of Fig. \ref{fig:HCFbothcuts} we see that a significant fraction of systems above $\beta_{\rm lim}$ merge, $20\%$ or more. At $\beta \gtrsim \beta_{\rm lim}$ the fraction peaks at $40\%$, with most mergers coming from systems that would otherwise be FAs. For $\beta \gtrsim 1$ some of all three outcomes contribute to the merger fraction, though the proportion of FAs that become Ms remains the highest, and about $10\%$ of systems that would disrupt on the first passage instead merge.

The stripe of low $r_{\rm min}$ can have potentially large effects on CBs at later passages (as we will detail in the next Section) since these systems evolve towards becoming Ds by subsequent passages being deeper and more prograde (up- and right-wards in these plots). However, now they must pass through this \textit{merger valley} to get there, with many likely being forced to merge before they can become Ds. The major exception to this will be systems that make large enough jumps in $\beta$ and $\cos(i)$ such that they pass right over it in one jump. 

\section{Multiple pericentre passages}\label{Multiplep}

\begin{figure*}
    \centering
    \includegraphics[width=\textwidth]{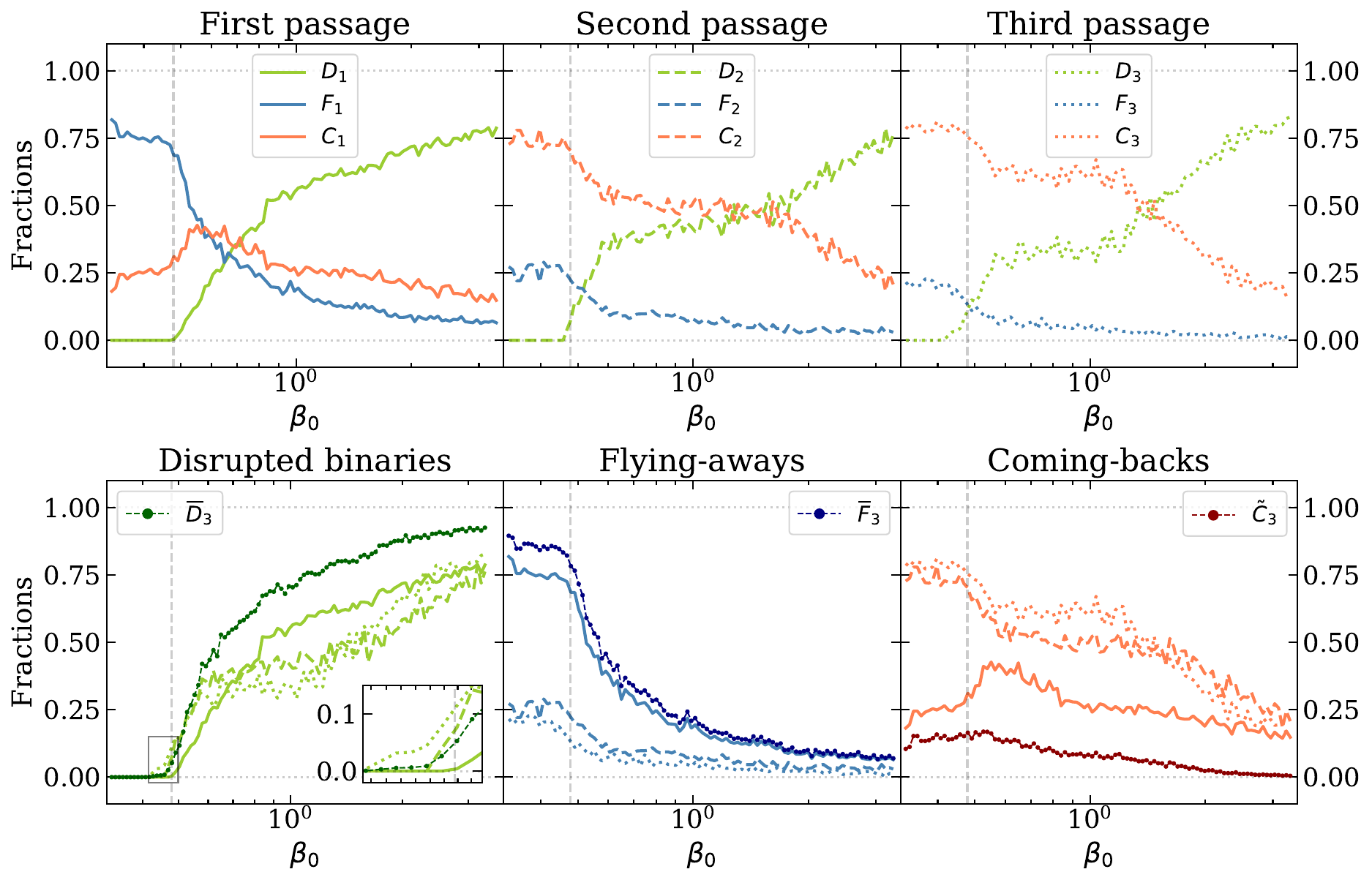}
    \caption{
    \emph{Top row}: 
    Fractions of Ds ($D_i$, green), FAs ($F_i$, blue) and CBs ($C_i$, coral) as a function of $\beta_0$ 
    at the end of the $i$-th passage. The vertical dashed grey line marks $\beta_{\rm lim}$.
    Panel 1 corresponds to passage 1 (bold lines), panel 2 to passage 2 (dashed lines), and panel 3 to passage 3 (dotted lines). 
     \emph{Bottom row}:  In each panel we show the fractions of systems ending up as Ds (left panel), FAs (centre panel) and CBs (right panel), respectively, after every passage (same lines as the top row). We add the overall fraction after three passages, weighted on the number of CBs from the previous passage (Ds in dark green, FAs in navy and CBs in dark red).}
    \label{fig:allfrac}
\end{figure*}

\begin{figure}
    \centering
    \includegraphics[width=\columnwidth]{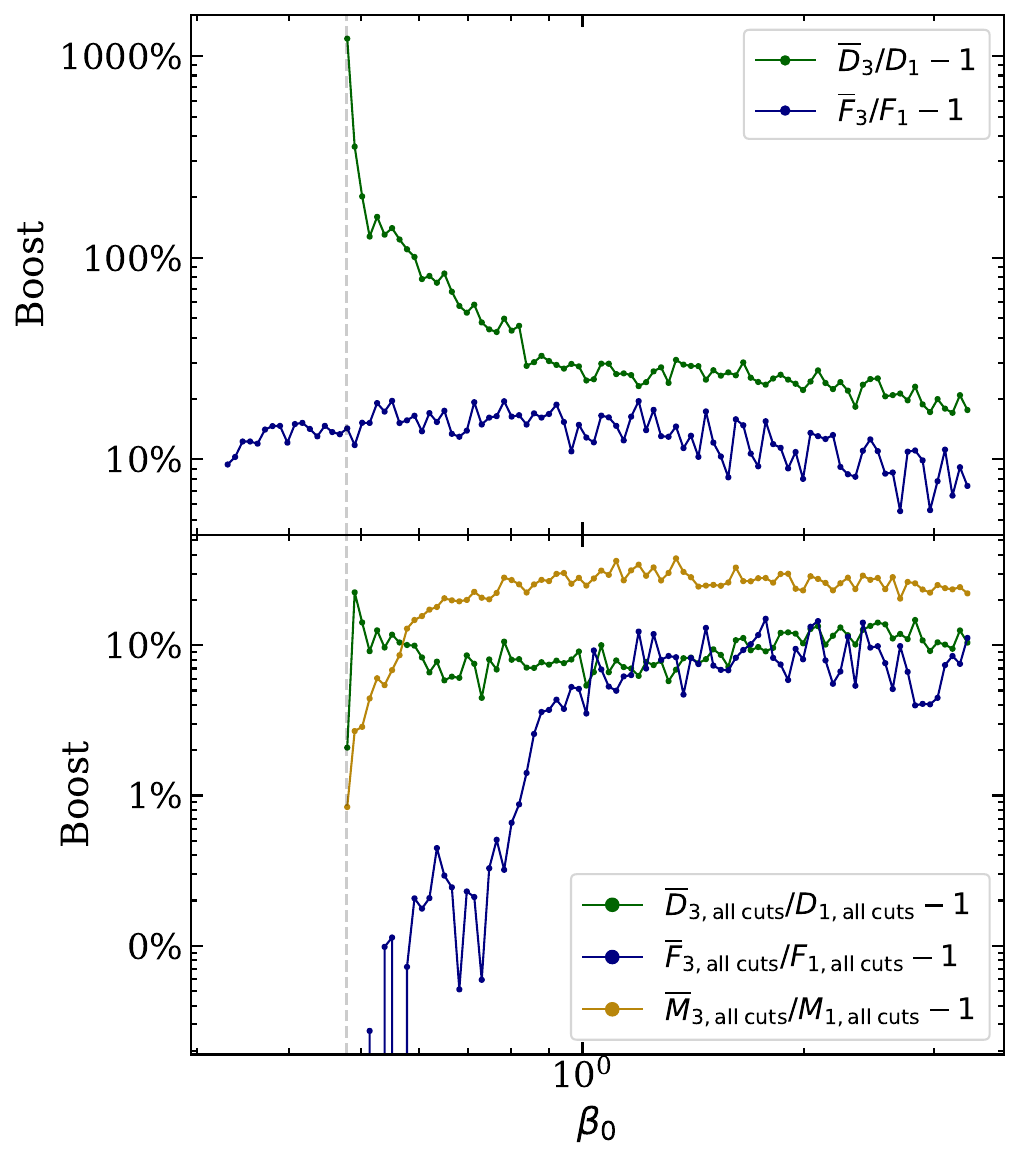}
    \caption{\emph{Upper panel}: Percentage boosts in the fractions of Ds (dark green) and FAs (navy), between the first and third pericentre passage, as a function of $\beta_0$.
    \emph{Bottom panel}: Same as above, but accounting for Ms (gold) and stellar ages for a $a_{\mathrm{b},0}=0.1$ AU, $m=4 M_\odot$, $q=\frac{1}{3}$ example system.}
    \label{fig:boosts}
\end{figure}

Motivated by the 
presence of a significant fraction of CBs, we proceed in this section with the analysis of multiple pericentre passages. We start by defining the initial conditions that allow us to follow the evolution of CB binaries from one passage to the next. We then compare the fractions of binaries in the three channels after three subsequent passages. 

We start from our first set of simulations considering circular binaries injected with different initial phases, inclinations, and orientations on a parabolic orbit with a diving factor $\beta_0$. We record their final state, separating them into Ds, FAs and CBs. The fates of any binaries that are disrupted or flyaway are resolved, but the CBs will undergo subsequent encounters and should (after sufficiently many passages) end up as either Ds or FAs.

Generally we assume that the final parameters of the binary after one passage tell us the initial parameters for the next passage, i.e. that any subsequent evolution away from pericentre or diffusion of the parameters is small. The one parameter we cannot safely do this with is the binary phase, $\phi$. It would be possible to calculate this (via the final binary phase and the binary and CM period) but as there will be many binary periods for a single CM period any small dispersion in the total change in phase will, when mapped to the interval $0<\phi<2 \pi$, lead to almost complete uncertainty on the actual phase. Thus for each comeback binary we simulate the next passage with $N_{n+1}=int(1/C_n)$ random phases (which means that for each encounter we simulate roughly the same number of systems). 

This procedure can be iterated for many passages, considering only the CBs from the previous passage as initial conditions for the next. Thus, each subsequent passage only explores the subspace of parameters where CBs occurred previously, and, unlike the first passage, all binaries will have $E_{\rm cm}<0$ and $e_{\rm b}>0$.

\subsection{Binary initial conditions for subsequent passages}\label{Subpassages}

We can denote the properties at the final properties at the end of the $n^{th}$ passage as, for example $e_{\mathrm{b},f}^n$. Similarly we can denote initial properties at the beginning of the next passage, for example $e_{\mathrm{b},i}^{n+1}$. We reserve the subscript $0$, for example $e_{\mathrm{b},0}$, for the initial conditions at the beginning of the first passage.

Many parameters are assumed to be unchanged from the end of one passage to the beginning of the next (i.e. $e_{\mathrm{b},i}^{n+1}=e_{\mathrm{b},f}^n$) but others need to be more carefully updated.

During an encounter, the binary changes its internal energy $\Delta E_{\rm b}$ and angular momentum $\Delta \mathbf{L}_{\rm b}$. We assume that total energy and angular momentum are conserved and thus these changes come at the expense of those of its CM orbit: 
\begin{align}\label{DeltaEL} E_{\mathrm{{\mathrm{cm}}}}^{n+1}&=E_{{\mathrm{cm}}}^n-\Delta E_{\mathrm{b},f},\\
    \mathbf{L}_{{\mathrm{cm}}}^{n+1}&=\mathbf{L}_{{\mathrm{cm}}}^n-\Delta \mathbf{L}_{\mathrm{b},f}.
\end{align}

From these we can derive the orbital parameters of the new orbit according to eqs. (\ref{ecm}) and (\ref{acm}) respectively as
\begin{align} 
e_{\mathrm{{\mathrm{cm}}}}^{n+1}&=  \sqrt{1+\frac{2E_{\mathrm{{\mathrm{cm}}}}^{n+1} \left(L_{\mathrm{{\mathrm{cm}}}}^{n+1}\right)^{2}}{G^2 M^2 m^3}}, \\
a_{\mathrm{{\mathrm{cm}}}}^{n+1}&=  -\frac{GmM}{2 E_{\mathrm{{\mathrm{cm}}}}^{n+1}}.\\ 
\end{align}
Apart from the phase, all binary orbital elements are unchanged from the end of the previous passage to the start of the next
\footnote{In theory, we should track also the orbital elements of the CM-trajectory ($i_{\rm cm}$, $\omega_{\rm cm}$ and $\Omega_{\rm cm}$) and reorient our inertial frame for the next passage. However, our assumption of a fixed CM trajectory means that our calculated $\mathbf{L}_{\rm cm}$ is inconsistent with our simulated $\mathbf{r}_{\rm cm}$ and $\mathbf{v}_{\rm cm}$ and thus $\omega_{\rm cm}$ and $\Omega_{\rm cm}$ are undefined. $i_{\rm cm}$ can be found (and is always small, as the CM trajectory changes only marginally between orbits) but given the lack of the other two angles we do not consistently adjust the frame of reference.}.


We have been working with lengths and times rescaled in terms of $\lambda$ and $\tau$ (see Section \ref{sec:units}). When analysing subsequent passages, these require adjustment. Some of the binaries that survive disruptions (CBs) will come back on elliptical CM-orbits, each with a different pericentre distance $r_{p}^{n+1}=a_{\mathrm{cm}}^{n+1}(1-e_{\mathrm{cm}}^{n+1})$. The masses and mass ratios are unchanged, and thus the total change to the characteristic units are captured by
\begin{equation}
\lambda^{n+1} = \frac{r_{\rm p}^{n+1}}{r_{\rm p}^{n}}\lambda^n = \frac{r_{\rm p}^{n+1}}{r_{\rm p}}\lambda
\end{equation}
and thus
\begin{equation}
\tau^{n+1} = \left(\frac{r_{\rm p}^{n+1}}{r_{\rm p}^{n}}\right)^\frac{3}{2}\tau^n = \left(\frac{r_{\rm p}^{n+1}}{r_{\rm p}}\right)^\frac{3}{2}\tau
\end{equation}
(where $r_{\rm p}$, $\lambda$ and $\tau$ are the values for the first passage).

The new diving factor is
\begin{equation}
\beta^{n+1} = \frac{r_{\rm t}^{n+1}}{r_{\rm p}^{n+1}} = \frac{a_{\mathrm{b},i}^{n+1}}{a_{\mathrm{b},i}^{n}}\frac{r_{\rm p}^{n}}{r_{\rm p}^{n+1}}\beta^n =  \frac{a_{\mathrm{b},i}^{n+1}}{a_{\mathrm{b},0}}\frac{r_{\rm p}}{r_{\rm p}^{n+1}}\beta_0
\end{equation}

Each subsequent passage effectively samples a subspace of the $\beta,  \cos(i), e_{\rm b}$ space (as shown in appendix \ref{app:ecc}) set by the CBs of the previous passage cluster. The CM-trajectory is only marginally perturbed ($r_{\rm p,n}\sim r_{\rm p}$ and $e_{\rm cm,n}\sim 1$) by the encounter and thus our previous analysis of the parabolic case is still representative of the behavior. The slight negative offset from $E_{\rm cm}=0$ biases the outcomes marginally towards CBs, and allows for rare cases where the binary disrupts and both single stars remain bound to the MBH (see \citealt{Kobayashi12} for further discussion).
 
We can denote the \textit{overall} fraction of CBs at the end of the $n$-th passage as 
\begin{equation}
\tilde{C}_{n}=\prod_{i=1}^{n}C_i,
\end{equation}
with $C_i$ the fraction of CBs for just the $i^{th}$ passage (it will also be useful to define $\tilde{C}_{0}\equiv 1$). We can then calculate $\bar{D}_{n}$, the overall fraction of Ds at the end of the $n$-th passages, by weighting the contributions at each passages based on the corresponding fraction of CBs giving 
\begin{equation}
    \bar{D}_{n}=\sum_{i=1}^{n}
\tilde{C}_{i-1}D_i.
\end{equation}\label{OverlineD}
Similarly for FAs, the overall fraction of binaries that fly away to populate the GC is 
\begin{equation}
\bar{F}_{n}=\sum_{i=1}^{n}
\tilde{C}_{i-1}F_i.    
\end{equation}

We choose to characterize the eventual fate based on the initial conditions at the first passage (especially $\beta_0$ and $i_0$), even when there may be multiple subsequent passages with varying initial parameters before the system resolves. This is equivalent to asking what the end state of a given sample of initial close encounters is, rather than focusing on the internal evolution between passages. In other words, if we start with a given binary we describe the state it ends in, agnostic to how many encounters it took to get there. 

\subsection{Results for three passages}

At each passage the remnant fraction of CBs decreases, eventually resolving into either FAs or Ds. The process could be considered complete when $\tilde{C}_n \rightarrow 0$. 
As we will show, most systems resolve after a few passages but there are regions of parameter space that produce persistent CBs and would require a large number of passages to asymptotically deplete. Thus we choose to show results after 3 passages, by which point the remnant fraction is small and the results likely capture all of the large scale behaviors of the high $n$ limit.

Returning to Fig. \ref{fig:hexbinHCF} we can now examine and compare the cumulative fractions after 3 passages (bottow row). The most striking feature is that there is now almost no dependence on inclination for the prograde and intermediate regime ($\cos(i)\gtrsim -\frac{1}{3}$. The population of CBs with intermediate inclination has almost completely resolved, splitting relatively cleanly into Ds for $\beta\gtrsim 0.6$ and FAs below that. There is a non-vanishing number of FAs even at high $\beta$, especially for retrograde binaries. Interestingly there are also some Ds with $\beta_0 < \beta_{\rm lim}$ and the fraction of disruptions close to that limit is markedly increased. Essentially all of the remnant CBs after 3 passages are retrograde, and the fraction is significantly reduced everywhere except for a cluster of strongly retrograde binaries.

We break this behavior down in Fig. \ref{fig:allfrac} where we show the fractions of the binary population that go into different channels at each passage. As CBs are more frequent in some regions of parameter space, and as the returning binaries are now commonly eccentric, we see different fractions from each passage. Most prominently the fraction of FAs decreases markedly, and the fraction of CBs increases - leading to diminshing returns on resolving the fate of CBs with each subsequent passage. The total remnant fraction after 3 passages, $\tilde{C}_3$ is generally small, at most 20\% for low $\beta_0$ and declining for deeper encounters. 

We also show the weighted sum of individual passages, $\bar{D}_3$ and $\bar{F}_3$, and can examine in particular how they are augmented from $D_1$ and $F_1$. Particularly we note that there are now many more disruptions at, and a small fraction slightly below $\beta_{\rm lim}$. It is generally expected that shallower encounters are more frequent \citep[see for example][]{StoneMetzger,Penoyre2025} and thus even a small boost to D at low $\beta$ may give a sizable increase to the actual number of disrupted systems.

We examine these boosts more directly in the upper panel of Fig. \ref{fig:boosts}, showing the percentage increase in the fraction of Ds and FAs after accounting for 3 passages. We see that FAs are boosted by around 10\% for all $\beta$, and that the effect for disruptions is higher still. For $\beta \gtrsim 1$ there is a consistent boost of around 20\% to the number of disrupted systems, and this grows significantly larger as we go to shallower encounters approaching $\beta_{\rm lim}$ (where very few systems disrupt on the first passage). Extending beyond 3 passages would (marginally) increase these boosts further, and thus we can conclude that accounting for CBs significantly increases the inferred number of FAs and, even more strongly, the number of Ds.

\subsubsection{Physical constraints for multiple passages}

As we did for a single passage in section \ref{sec:phys_constraints} we can consider limitations to which systems can comeback based on the physical time and length scales of the binary. Using again our example system, with $a_{\mathrm{b},0}=0.1$ AU and $m=4 M_\odot$, we can discard the effects of CBs whose CM trajectory is longer than their expected stellar lifetime (\textit{period cut}) or where the binary members come close enough to tidally interact (\textit{merger cut}). CBs which fail the period cut will be treated as FAs, whilst those which fail the merger cut are classified by a separate category of merged systems, Ms. Including multiple passages means that now both cuts can effect the total number of Ds (whearas for a single passage the period cut only shifts the balance between CBs and FAs) as some systems previously analyzed will now ``fail'' to comeback.

\begin{figure}
    \centering
    \includegraphics[width=\columnwidth]{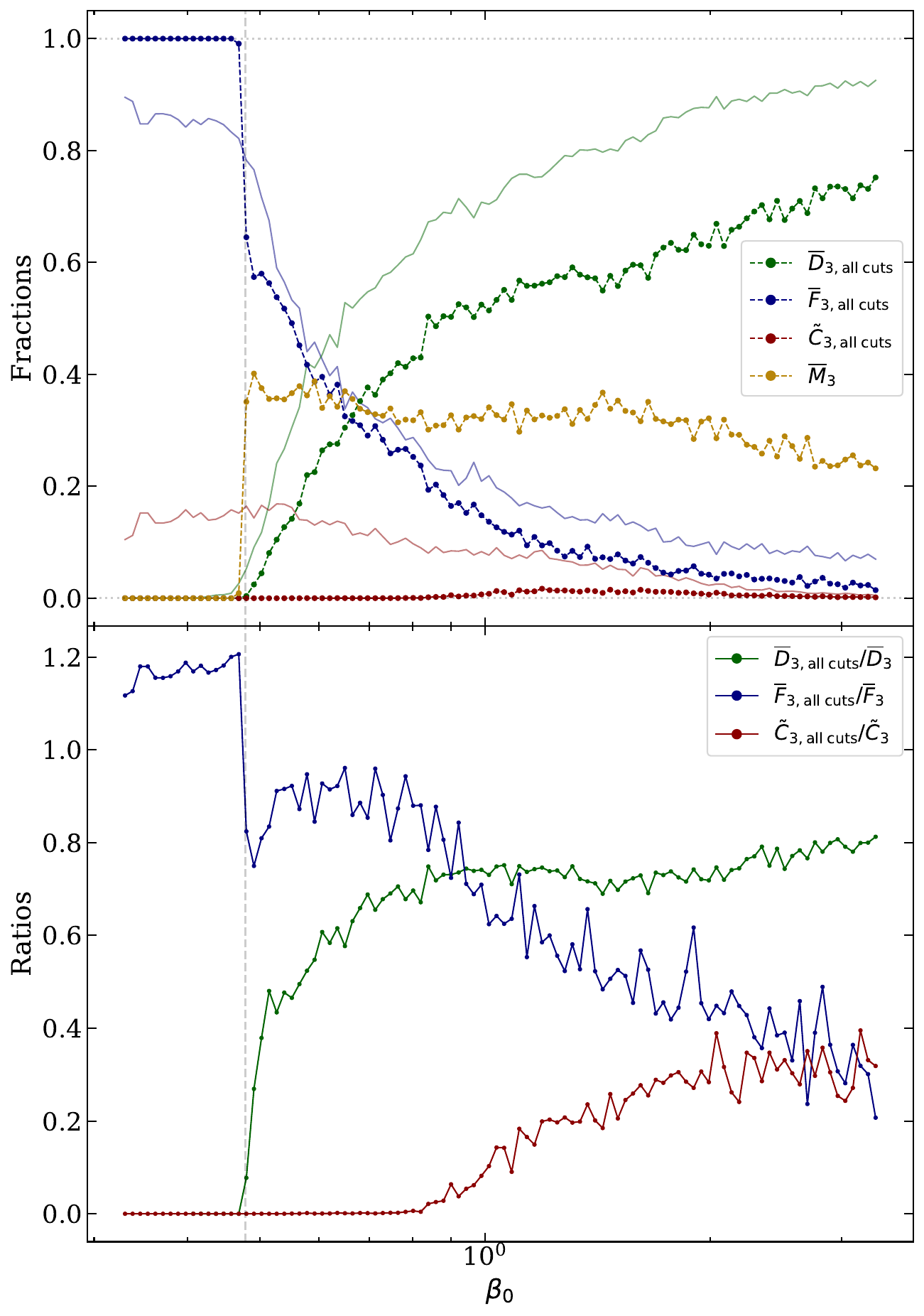}
    \caption{\emph{Upper panel}: Fraction of Ds, FAs, CBs and Ms after 3 passages (dotted lines - green, blue, red and yellow respectively) including period and merger cuts based on our example binary. We compare these to the fraction of Ds, FA and CBs without cuts (thin lines). 
    \emph{Lower panel}: Ratios between the fractions of Ds, CBs, FAs (same colour scheme), at the end of three encounters, before and after accounting for mergers and stellar age.}
    \label{fig:nomerg-merg-p3}
\end{figure}

Fig. \ref{fig:nomerg-merg-p3} shows effects of these physical constraints on the fractions of Ds, FAs, CBs and Ms after three passages. As we saw for the first passage all low $\beta$ systems are FAs, as the CM period of CBs is so large here and the stars expire before they can return. Above $\beta_{\lim}$ Ms are common ($>20 \%$ of outcomes), and the net effect of these is to primarily suppress CBs and Ds. In particular CBs that, after the first passage, would likely have evolved towards disruption, are now pushed towards merging. It is still the case that Ds are ubiquitous, if reduced in number by around $20\%$ for $\beta \gtrsim 1$, though almost no systems close to $\beta_{\rm lim}$ disrupt. The number of FAs above $\beta_{\rm lim}$ is decreases with increasing $\beta$, and may become negligible for slightly higher $\beta$ than simulated here. These results would suggest (if our example system is representative of the broader population) a significant number of merged binaries in the Galactic Center, which are an interesting object of study in their own right as they could possibly explain G-type objects \citep[see][]{CiurloNaoz2020,StephanNaoz}.

The bottom panel of Fig. \ref{fig:boosts} shows the percentage boosts, subject to these cuts, to Ds and FAs when including 3 passages. The cuts slightly reduce the importance of subsequent passages, as they generally reduce the number of CBs, but there is still a $\sim 10 \%$ or greater increase to the number of disrupted systems, and slightly less than $10\%$ for FAs with $\beta_0>1$. The boost to FAs goes to zero now at low $\beta$ as no CBs in this space survive the period cut. The boost to Ds is consistently of order $10\%$, and the boost to FAs is similar for $\beta\gtrsim 1$. The number of Ms is significantly boosted when including 3 passages, by $20\%$ or more for $\beta\gtrsim 0.6$. Thus we see again that if we did not follow the full evolution of CBs to their resolution we would moderately underestimate the fraction of Ds, FAs and Ms that are produced by the Hills mechanism.

\section{Closer look at ejections and captures}
\label{Results}
As seen we have shown in section \ref{Multiplep},  multiple pericentre passages should be taken into account to fully characterize the distributions of ejected and captured stars as well as of FAs. Thus, in this Section, we present the characteristic distributions of velocities, semi-major axes and inclinations of Ds and FAs following 3 passages.
The characteristic units (as derived in Appendix \ref{app:units}), which can be rescaled for any three bodies that satisfy the assumptions of the Hills mechanism, are:

\begin{itemize}
    \item $\nu_{\rm D}=Q^\frac{1}{6}\sqrt{2\frac{Gm_{\rm cap}}{a_{\mathrm{b},0}}}$ is the characteristic velocity of an ejected star, where $m_{\rm cap}$ is the mass of the corresponding captured component (showing that we can get substantially faster ejections for the lighter component);
    \item $\alpha_{\rm D}=\frac{1}{2}Q^\frac{2}{3} \frac{m}{m_{\rm ej}}a_{\rm b0}$ is the characteristic semi-major axis of the captured component of a disrupted system, where $m_{\rm ej}$ is the mass of the other component, and $\delta_{\rm D}=2Q^{-\frac{1}{3}} \frac{m_{\rm ej}}{m}\beta^{-1}$ is the magnitude of the characteristic change in the CM trajectory eccentricity (from the initial value of 1 for a parabolic orbit).
    \item $\nu_{\rm b}=\sqrt{\frac{Gm_1m_2}{ ma_{\rm b,0}}}$ is the characteristic ejection velocity of an ejected binary;
    \item $\delta_{\rm b}=Q^{-\frac{2}{3}}\frac{m_1m_2}{m^2}\beta^{-1}$ is the characteristic deviation from 1 of the CM-eccentricity of an ejected binary;
\end{itemize}
where for our example binary with initial semi-major axis of $0.1~\rm AU$, $m=4M_{\odot}$, $q=\frac{1}{3}$ we find $\nu_{\rm D}\approx 1300 \ \mathrm{ km \; s^{-1}}$, $\alpha_{\rm D} \approx 6.67 \cdot 10^{2} \ \mathrm{AU}$ and $\delta_{\rm b} \approx 0.015 \beta^{-1}$, assuming the lighter component is ejected 
and $\nu_{\rm b} \approx 82 \ \mathrm{ km \; s^{-1}} $ and $\delta_{\rm b}\approx 1.88\cdot 10^{-5}\beta^{-1}$ if the binary survives the encounter. Note that only $\delta_{\rm b/D}$ depend on $\beta$, generally deeper encounters do not generally produce more extreme outcomes.

\subsection{Disrupted binaries}
\label{sec:test}
 In this section, we present our results for the star properties following a binary tidal separation: distributions of velocity for the ejected stars, and those of eccentricity and semi-major axis for the captured stars. We show results obtained after three passages and compare them with those obtained after the first encounter.
\subsubsection{Ejected stars}
Fig. \ref{fig:allfrac} shows the properties the ejected stars, as a function of $\beta_0$, after one and three passages. We see that the characteristic $\nu_D$ well captures the measured ejection velocity, except at low $\beta$ where it is a slight overestimate. The distribution after 1 passage shows that for $\beta \gtrsim 0.6$ the majority of systems have $v_{\rm ej}$ between $0.75 \nu_{\rm D}$ and $1.25 \nu_{\rm D}$. 

\begin{figure}
    \centering
    \includegraphics[width=\columnwidth]{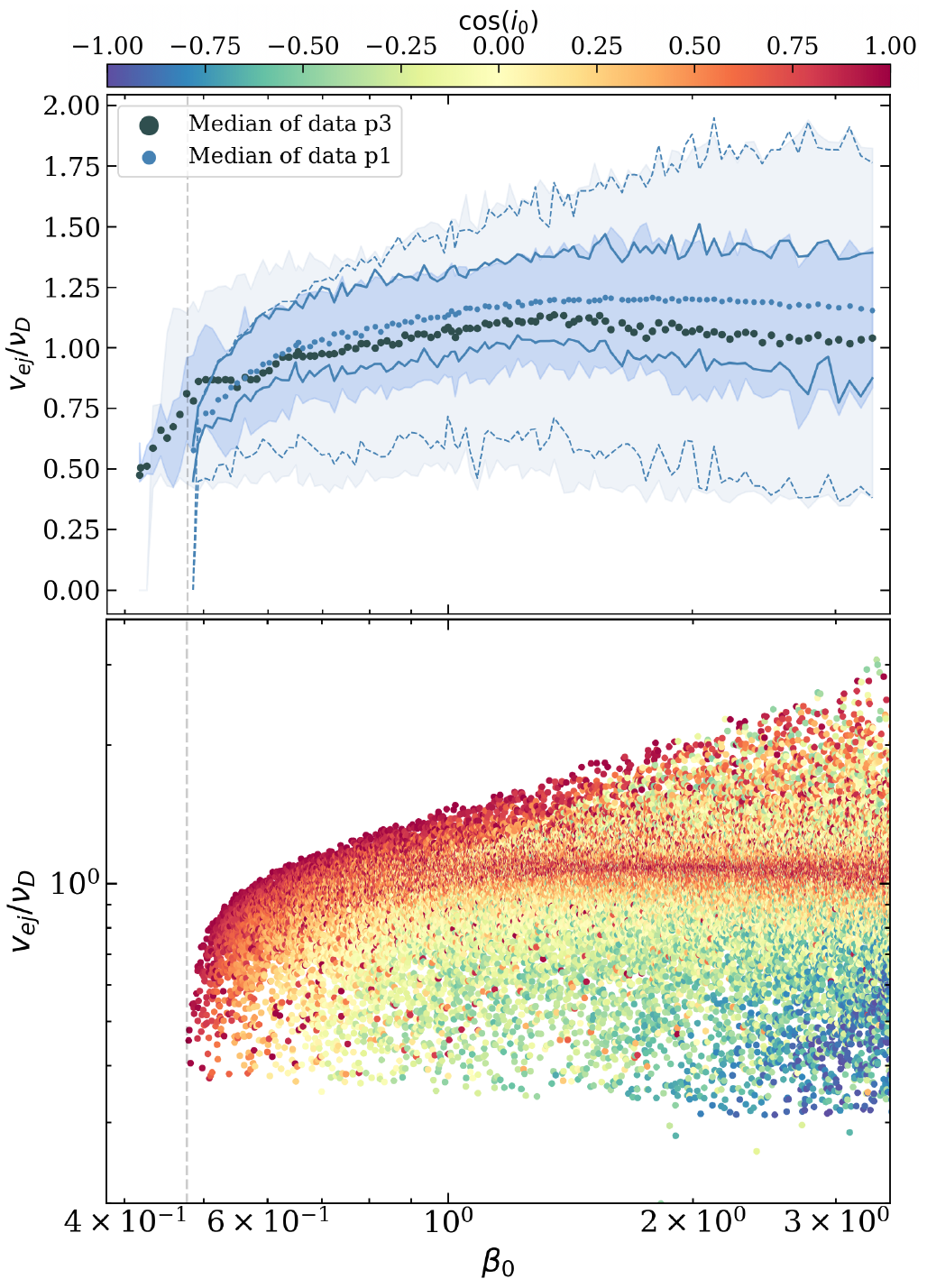}
    \caption{Distributions of the velocities of ejected stars as function of $\beta_0$. Ejection velocities are expressed in units of the characteristic velocity $\nu_{\rm D} $ (e.g. $\approx 1300~\mathrm{km\;s^{-1}}$ for our binary of choice). \emph{Upper panel}: Distribution after one passage (light blue) and three passages (dark blue). The solid and dashed lines show the $68\%$ and $95\%$ confidence intervals for the first passage. The shaded regions show the equivalent for 3 passages.
    \emph{Lower panel}: Distributions of ejection velocities after one passage as a function of $\beta_0$, coloured by $\cos(i_0)$), for 100,000 sets of uniformly-sampled initial conditions.
}
   \label{cdf_hvs_code}
\end{figure}

As we saw earlier, CBs increase the number of Ds produced in shallow encounters, and in particular they extend their production below $\beta_{0,\text{lim}}$ (see Fig. \ref{fig:allfrac}). These additional ejected stars have median velocities {\it higher} than those produced in the first passage for$\beta_{0} \approx \beta_{\rm lim}$, while they substantially extend the high-velocity tail at each $\beta_{0}$ for $\beta_{0} \le 0.75$ (top panel in Fig.\ref{cdf_hvs_code}). Going towards deeper encounters the median velocity after three passages is slightly lower than after one encounter. I.e. the extra disruptions fro later passages are generally of lower velocity (though note that including more passages only adds systems, all of the high velocity ejecta are still present)

We present the role of inclinations in the bottom panel of Fig. \ref{cdf_hvs_code}, showing only our results for the first passage for clarity. Although both prograde and less inclined ($i_0<\pi/2$) binaries can reach high ejection velocities, there is an overall trend at a fixed $\beta_0$, where increasingly prograde binary progenitors produce increasingly faster ejections. In this trend, three regions stand out. At shallow encounters, the high-velocity tail is completely dominated by prograde binaries, which also feature prominently at the value of the median velocity $v_{\rm ej}/[\nu_{\rm D}]\simeq 1$ for all $\beta_0$. On the other hand, the low-velocity tail for our deepest encounters is entirely due to retrograde binaries. 
Multiple encounters do not affect the maximum ejection velocities but add ejections, of mostly retrograde binaries, across the full velocity range, which is broadened at lower velocities with respect to the first passage.

\subsubsection{Captured stars}
\begin{figure}
    \centering
    \includegraphics[width=\columnwidth]{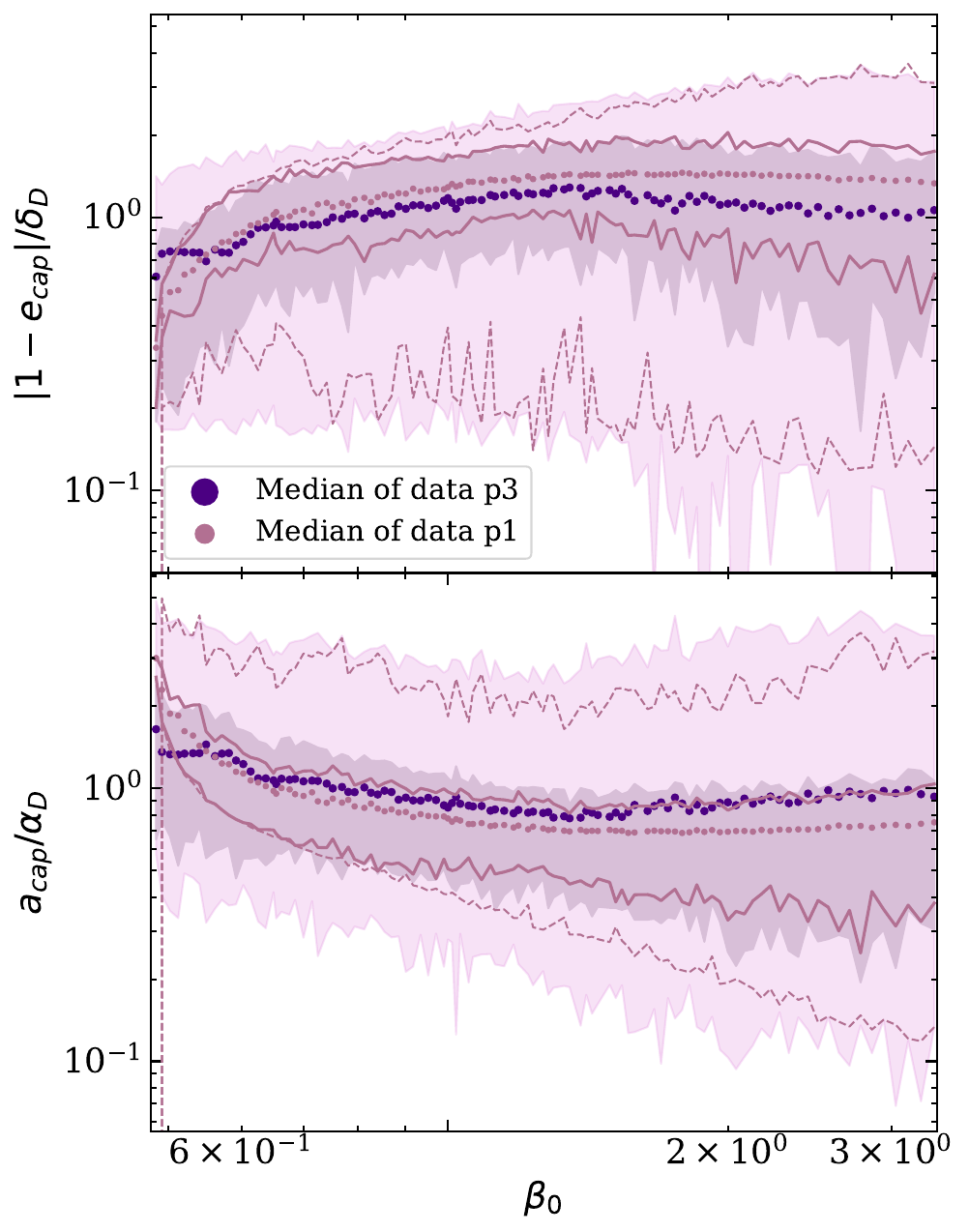}
    \caption{Similar to Fig. \ref{cdf_hvs_code} we show the properties of the captured star from a disrupted binary. We show the distributions of the eccentricities (upper panel) and semi-major axes (lower panel) of captured stars as a function of $\beta_0$.
    $|1-e_{\mathrm{cap}}|$ is expressed in units of $\delta_{\rm D}$ (e.g. $\approx0.015\beta^{-1}$ for our example binary) and 
    $\alpha_{\mathrm{cap}}|$ in units of $\alpha_{\rm D}$ (e.g. $\approx 667$ AU for our example binary).}
    \label{cdf_sstar}
\end{figure}

Fig. \ref{cdf_sstar} shows the distribution of the characteristic changes (from 1) in eccentricity and of the semi-major axes of the S-stars' orbits around the MBH. Analogously to those of the ejected stars, both are altered at shallow encounters by additional disruptions from CB binaries. With respect to the change in eccentricity, the median is about the characteristic scale $\delta_D$, such that $e_{\mathrm{cap}}\sim 0.98$ for our example binary. High eccentricities for the bound stars are compatible with the observed eccentricities in the S star cluster (for instance S2 has an eccentricity of $0.88466\pm 0.00018$). As regards the semi-major axis, the median sits around the characteristic semi-major axis, which for our chosen binary is of the order of $10^2$AU, also compatible with the observed values in the S star cluster (S2 has a pericentre distance of about 120 AU and semi-major axis of about 970 AU). However, the dispersion is large especially for deeper encounter, spanning one and half order of magnitude.

\subsection{Hypervelocity stars fraction}
In this section, we use our example binary to get an estimate for the fraction of encounters that leads to Hyper Velocity Stars (HVSs). These are stars ejected from the GC with velocities greater than the escape speed from the Galaxy, and can therefore be potentially observed on their way out in the halo of our Galaxy. As a velocity threshold, we choose $v_{\rm esc} = 1000$ km $\rm s^{-1}$ which is a typical galactic escape speed for a McMillan model of the MW potential \citep[see][]{McMillan2017} and
is a reasonable assumption in our case, considering the order of magnitude of the characteristic velocity $\nu_{\rm D}$ for a solar mass binary with initial separation of 0.1 AU.
We define
\begin{equation}
    H_{j} =P(v_{\rm ej}>v_{\rm esc}|D_j),
\end{equation}
the fraction of systems in passage with $v_{\rm ej}>v_{\rm esc}$ and thus the overall fraction after multiple passage is
\begin{equation}
    \bar{H}_{n}=\sum_{j=1}^{n}
\tilde{C}_{j-1}H_j.
\end{equation}\label{OverlineH}

\begin{figure}
    \centering
    \includegraphics[width=\columnwidth]{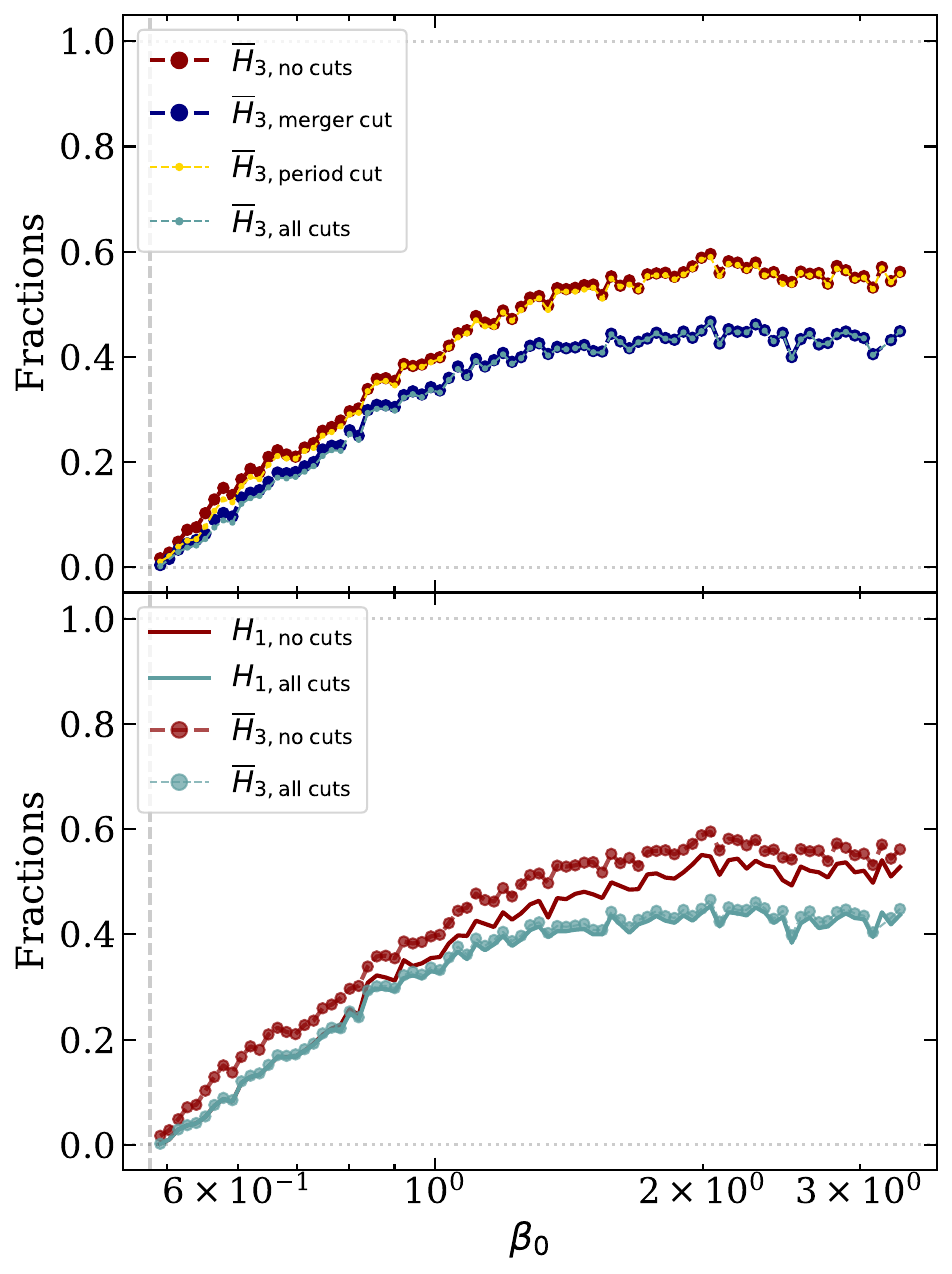}
    \caption{
    \emph{Top panel}: fractions of HVSs (unbound stars with $\rm v_{\rm ej}>1000$ km s$^{-1}$) ejected after three passages, using our example binary, as a function of $\beta_0$. Before any cuts (dark red), after period cut ($\rm P_{cut}=10^{8} yr$) (yellow), after merger cut (dark blue) and after applying both cuts at the same time (light blue). 
    \emph{Bottom panel:} fractions of HVSs after the first and third passage (solid and dotted lines, respectively) without any cuts (dark red) and with both cuts applied together (light blue).}
    \label{fig:HVsGC}
\end{figure}

Fig. \ref{fig:HVsGC} shows the fraction of interactions, as function of $\beta$, that result in our example binary disrupting and giving a HVS. We include both the period and merger cut, the former of which has only a marginal effect except at low $\beta$, whilst the latter reduces the fraction by around a quarter. We also show the difference between the fractions after one and three passages, where we can see that without the period and merger cut later passages significantly boost the fraction of HVS, but that this boost is much reduced when including the cuts. Thus for a system like our example binary most HVSs are produced on the first passage. We can understand this as a consequence of the fact that Ds at later passages tend to have lower $v_{\rm ej}$ generally. \citet{Sari10} showed that both stars, regardless of $q$, are equally likely to be captured/ejected, however the characteristic velocity depends on the mass of the captured star. This means that in the majority of these HVSs are the lighter component, and that the heavier component can only become a HVS for $\beta\gtrsim 0.6$. For $\beta \gtrsim 1$ we find that more than $40\%$ of encounters involving our example binary produce a HVS.

\subsection{Flying-away binaries}
\begin{figure}
    \centering
    \includegraphics[width=\columnwidth]{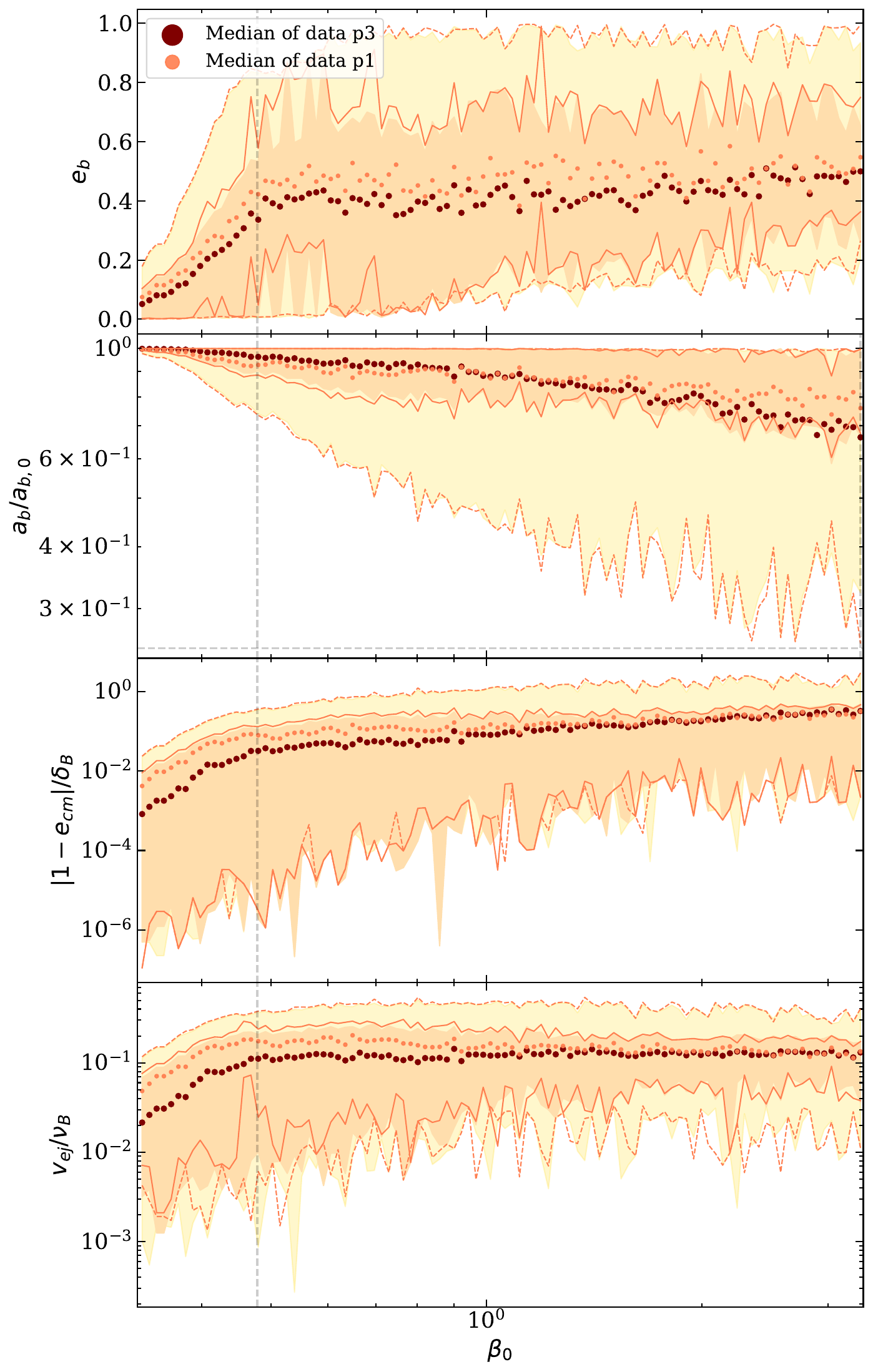}
    \caption{Similar to Fig. \ref{cdf_hvs_code} and \ref{cdf_sstar} we show the properties of a fly-away binary, after one passage (coral) and three passages (dark red).
    From top to bottom we show the: binary eccentricity, relative semi-major axis, deviation from 1 of the eccentricity of the CM trajectory in units of $\delta_B$ (e.g. $\approx 1.88\times 10^{-5}$ for $\beta_0=1$), and ejection velocity in units of $\nu_{\rm B}$ (e.g. $\approx 82 ~\mathrm{km\;s^{-1}}$ for our binary of choice).}
    \label{cdf_fas}
\end{figure}

In Fig. \ref{cdf_fas} we show the properties of FAs, binaries that are unbound from the MBH but likely remain in the Galactic Center and are thus another potential signature of Hills mechanism interactions. Although their progenitors are circular, FAs are {\it eccentric} binaries with a typical eccentricity of $ e_{\rm b} \sim 0.5$ (approaching the average for a thermal distribution of eccentricities of $\frac{2}{3}$) for $\beta_0\gtrsim\beta_{\rm lim}$. Some near-circular binaries persist up to $\beta \sim 1$, and below $\beta_{\rm lim}$ the induced eccentricity reduces. The semi-major axis of the binary is slightly reduced by the interaction, but by a factor of less than $\frac{1}{2}$ even at the highest $\beta$. Even for shallow interactions, $\beta \sim 0.3$, many binaries have still been imparted with eccentricities of order $e_{\rm b} \sim 0.1$, whilst their semi-major axes are essentially unchanged. This suggests that the Hills mechanism is efficient in transferring angular momentum even when energy transfer is marginal.

The change in eccentricity of the CM trajectory and ejection velocities are low, around an order of magnitude less than the limiting  characteristic units $\delta_{\rm B}$ and $\nu_{\rm B}$. For our example binary these correspond to $\Delta e_{\rm cm} \sim - 10^{-6}$ and $v_{\rm ej}\sim 10 \mathrm{km \; s^{-1}}$ (two or three orders of magnitude slower than stars ejected by disruptions), giving a population of binaries only marginally unbound from the MBH on very radial orbits.

\label{sec:like}

\section{Discussion and conclusion}
\label{Concl}
We have investigated multiple close encounters between a stellar binary and a MBH, with the aim of characterizing the properties of the resulting population of binaries, single stars and merger products.
The mathematical formalism underlying our investigation is based on the analytical treatment of the Hills mechanism \citep{Hills1988}. We model the interaction as a restricted three-body problem under approximations made possible by exploiting the very large difference in mass and length scales between the binary and the MBH \citep[following][]{Sari10,Brown18,Kobayashi12}. We also tested our results against a full direct three-body integration and confirm that our results are accurate.

We started with a population of circular binaries on parabolic trajectories around the MBH, for a random distribution of binary phases, orientations, inclinations and diving factors. We first analysed the fates of our systems after a single tidal encounter. Systems can resolve into two definitive channels --disruptions (dubbed {\it Ds}) and ejections of the binaries (Fly-Aways {\it FAs}). There is a thid temporary channel --the binaries survive the encounter and come back ({\it CBs}) to the MBH for a subsequent encounter. By specifying the finite sizes of the stars another resolution, mergers ({\it Ms}), are also possible (see, Section \ref{Results}). Finally, employing the complete framework proposed in \citep[][]{Kobayashi12}, we map the initial population into the various fates mentioned above, and analyse each of them after three encounters. 

In the following, we summarise our results. We quote fractions marginalizing over all parameters (see Table \ref{tab:table}) and present both the results of a single encounter --which depend on the properties of the incoming binaries (diving factor $\beta$, eccentricity $e_{\rm b}$ and inclination $\cos{i}$)-- and results at the end of multiple encounters, cast in terms of the initial parameters (i.e. of those at the beginning of the {\it first passage}), such as the initial $\beta_0$ and inclination $\cos{i_0}$, of the binary. 
\begin{itemize}
\item{\bf CBs.} There is a significant fraction of CBs ($\sim 26\%$ ) after the first passage; this fraction is progressively reduced in later passages, and reaches $\sim 8\%$ at the end of the third one, as CBs become FAs or Ds. As a consequence of the first encounter, when approaching as circular binaries, CBs become eccentric, and follow deeper-diving CM-trajectories than initially; if they change their inclination at all, they become more intermediate or prograde (see, Fig. \ref{fig:propC1}). CB binaries preferentially originate from {\it mildly retrograde} and {\it retrograde} encounters, that can occur through the whole $\beta$ range explored here, with eccentricity up to $e_{\rm b}\approx 0.8$; approaching {\it prograde} binaries can also become CBs,  when on shallow encounters ($\beta \lesssim 0.48$) and modest or high eccentricities of at least $e_{\rm b}\approx 0.2$ (see, rightmost upper panel of Fig.~\ref{fig:hexbinHCF} and rightmost column of Fig. \ref{hexbinHCF_ecc}). 
\item{\bf Ds.}  Binary disruption is the most common outcome in our parameter space. The disruption fraction is $\sim 46\%$ after one passage, and increases to $\sim 60\%$ (with an almost constant boost of 16\% for $\beta_0>1$) at the end of the third encounter, due to the fresh batch of CBs, whose aforementioned properties (eccentric, more prograde, on deeper encounters) make them more prone to disruption in the subsequent passage. Indeed, binaries end up preferentially disrupted for \emph{intermediate} and \emph{prograde} orbits and for deep encounters ($\beta\gtrsim0.4$); retrograde binaries can also be disrupted, if highly eccentric ($e_{\rm b} \gtrsim 0.7$), and/or on a deep encounter, increasingly deeper for increasingly retrograde binaries (see leftmost upper panel of Fig.~\ref{fig:hexbinHCF} and leftmost column of Fig. \ref{hexbinHCF_ecc}). This explains why, in the parameter space of the initial encounter (leftmost lower panel of Fig.~\ref{fig:hexbinHCF}),  multiple passages allow disruptions to populate the space below $\beta_{\rm 0} \approx 0.48$ as well as retrograde inclinations (i.e. $\cos{i_0}\lesssim -0.5)$, where instead no disruptions occur in the first passage.
\item{\bf Captured stars.} The binary member captured by the MBH, ends up on a bound eccentric orbit; when using our example binary, the median values of its orbital properties are the typical ones: the eccentricity is $e_S\sim 0.98$ and the semi-major axis is $a_S\approx 6.67\times 10^2 AU$ (with the bigger spread, between $\approx67$AU and $\approx 2000$AU, at deeper encounters), consistent with the orbit of S2 in the GC, (see Figure \ref{cdf_sstar}). 
\item{\bf Ejected stars.} After three passages, ejected stars have a median velocity about the characteristic velocity $\nu_{D} = Q^{1/6}\sqrt{\frac{2Gm\mu}{m_{12}a_{\mathrm{b},0}}}$ ($\approx 1300\rm\;km\;s^{-1}$ for our example binary) for $\beta_0 \gtrsim 0.48$, while lower velocities down to $\approx 0.5 \nu_{D}$ are reached for smaller $\beta_0$s. The velocity distribution is shaped mostly by the first encounter, while subsequent passages extend the distribution at lower $\beta_0$s, and lower the median velocity at deep encounters by $\approx0.1\nu_{\rm D}$ : e.g. for our example binary the median after three passages at deep encounters is smaller by $ \approx130\rm\;km\;s^{-1}$ than that at first passage, (see Fig. \ref{cdf_hvs_code}).
\item{\bf HVSs.} When considering our example binary, we refer to HVSs as stars ejected with a velocity greater than $1000$\; km s$^{-1}$, which occur preferentially (i.e. for more than 50\% of the total) for $\beta_0\gtrsim 1$, with the dominant contribution due to the first passage. The fraction of HVSs after one passage is about $\sim 31\%$ of the total injected binaries, reaching $\sim 35\%$ at the end of the third one. Preferencially, the lighter compation is ejected as an HVSs independenty of $\beta$s: this is because CBs are on bound orbits, and the more massive --carrying most of the orbital energy-- is more likely to continue on a bound orbit, i.e. to be ``captured'', \citep[see][]{Kobayashi12}.
\item{\bf FAs.} The FA fraction goes from $\sim 28\%$ at the end of the first passage to $\sim32\%$ at the end of the third one (with the an almost constant $10\%$ boost below $\beta_{\rm lim}$, which then decreases at deeper encounters). FA binaries preferentially originate from shallow encounters (below $\beta_0\approx 0.48$); additionally, when the approaching binary has an intermediate or prograde inclination, it becomes a FA if of low eccentricity up to $e_{\rm b}\approx 0.4$, while the opposite is true for retrograde binaries (see, central upper panel of Fig.~\ref{fig:hexbinHCF} and central column of Fig. \ref{hexbinHCF_ecc}). Their fraction distribution is shaped primarily by the first passage because, once again, the CBs are more inclined to disrupt at the subsequent encounter (due to their orbital properties). FAs are ejected on slightly hyperbolic orbits ($e_{\mathrm{cm}}\gtrsim 1$), with velocities lower than those of ejected stars by $\sim$ 2 order of magnitudes ($\approx 8.2\;\rm km \;s^{-1}$ for our example binary). They get tighter after tidally interacting with the MBH (due to energy conservation) and become eccentric: $e_{\rm b}\lesssim 0.4$ below $\beta\approx 0.48$ and  $e_{\rm b}\lesssim 0.6$ for deeper encounters (see Fig.~\ref{cdf_fas}).
\item{\bf Impact of lifetime.} When considering our example binary, we take into account the finite lifetime of MS stars to assess if CBs in fact have time to go through a subsequent encounter. CBs are reduced by a $\sim 12 \%$  (to $\sim 14\%$) at first passage.
This progressive reduction in the CB fraction, by the end of the third encounter determines a reduction of the D-fraction by $\approx 3\%$ and a boost of the FA-fraction by $\approx 9\%$.
\item{\bf Mergers.} When considering our example binary, we can introduce mergers as a fourth channel. 
By the end of the third passage, the fraction of mergers is $\approx 31 \%$ and it causes the D-fraction to be reduced by $16\%$, the F-fraction by by $10\%$ and the CB-leftover fraction by $4\%$. 
If we consider, on top of mergers, the effect of the stars' life time, after three pericentre passages $43.14\%$ of the binaries disrupts, $29.84\%$ flies away and $26.54\%$ merges (with a leftover of CBs of $0.47\%$).
\item The median parameters distributions after multiple encounters is generally similar (slightly more moderate) to the distribution after the first passage. The exception to this is shallow encounters (low $\beta$) where Ds from later passages can result in faster ejecta and tighter bound captured stars.
\item Including three passages boosts the number of Ds by $20\%$ or more, and markedly increases the number of disruptions from shallower initial encounters. FAs are boosted by about $10\%$, showing that CBs after the first passage are more likely to resolve as Ds. For stars of finite size and lifetime we can also apply a period and merger cut, reducing the boost to Ds to about $10\%$, but with $20\%$ or more systems merging after 3 passages compared to just one. We limit ourselves to 3 passages to limit computational cost, with most CBs having been depleted by this point, but simulating further passages is possible and will further (marginally) boost the number of Ds, FAs and Ms.
\end{itemize}

\begin{table}
  \centering
\resizebox{\columnwidth}{!}{%
\begin{tabular}{|c|*{6}{c|}}
\hline
\textbf{} & \textbf{} &\textbf{Ds}\% & \textbf{FAs}\% & \textbf{CBs}\% & \textbf{Ms}\% & \textbf{HVSs}\%\\
\hline
\multirow{4}{*}{Passage 1} 
& No cuts & 45.63 & 28.37 & 26 & - & 31.06 \\
& Lifetime & 45.63 & 40.05 & 14.32 & - & 31.06 \\
& Mergers & 39.40 & 18.86 & 20.13 & 21.61 & 26.86 \\
& Mergers and lifetime & 39.40 & 29.56 & 9.44 & 21.61 & 26.86 \\
\hline
\multirow{4}{*}{Passage 2} 
& No cuts & 55.41 & 31.19 & 13.40 & - & 33.85 \\
& Lifetime &54.36 & 40.92 &4.72 & - & 40.44 \\
& Mergers & 42.85 &20.67 &8.11 &28.36 & 33.32 \\
& Mergers and lifetime & 42.46 & 29.81& 1.95 & 25.78 & 32.99 \\
\hline
\multirow{4}{*}{Passage 3} 
& No cuts & 59.61 & 32.32 & 8.07 & - & 34.72 \\
& Lifetime & 57.26 & 41.07 & 1.67 & - & 34.15 \\
& Mergers & 43.85 & 21.39 &4.05 & 30.70 & 27.34 \\
& Mergers and lifetime &43.14 & 29.84 & 0.47 & 26.54 & 26.98 \\
\hline
\end{tabular}
} 
\caption{Fractions of the total initial systems ending up in each of the channels at different passages and considering different or no cuts. Fractions are here marginalized over all the parameters, including the diving factor.}
 \label{tab:table}
\end{table}

\cite{StephanNaoz2016} find a significant fraction of mergers as well, of about 13\% of their initial population after a few million years and 29\% after a few billion years. However, the origin of the mergers is different: what we find are dynamical mergers, occuring on the binary timescale at pericentre, while they investigate secular Kozai-Lidov processes. 
On the other hand, our findings are in agreement with
\cite{AntoniniFaber2010}, who also follow binaries that remain bound for several revolutions around the SMBH with N-body simulations. They find that HVSs are primarily produced in the first passage while collisions and mergers increase significantly for multiple encounters (due to Kozai-Lidov resonance of the internal binary).
\citep{MandelLevin2015} and \cite{BradnickMandelLevin2017} estimated that the fractions of mergers for a population of 1000 binaries in radial and shallow encounters ($\beta_0\approx 0.5$), is, respectively, $\approx6\%$ and $80\%$. They follow the binaries until their complete depletion (into HVSs or mergers, without analysing FAs). 
Our work complements the above results by exploring the full $\beta$ and $\cos(i)$ range, especially as the treatment used allows many quick and efficient simulations. We follow systems to their final outcome and show the properties of the resulting systems.

Most of our findings are general, allowing any choice of the initial binary. These results apply to any system where the physical length and timescales do not interrupt the repeated interactions, and is thus directly relevant for a tight compact object binary (a promising progentior population for EMRIs). We also choose a specific example system, a massive stellar binary, aiming to highlight the relevance of this approach to HVS candidates and the nuances presented by a short-lived system that may be subject to the binary merging. 
In a follow-up paper, we will further explore different astrophysically motivated initial binary populations, to provide valuable predictions and insights on a broad range of transient phenomena (EMRIs, TDEs, QPEs) occurring in the GC, and the impacts on its stellar population. 
-------------------
\section*{Data Availability}
The data underlying this article will be shared on reasonable request to the corresponding author.

\section*{Acknowledgements}
The authors thank Niccolò Veronesi and Sill Verberne for helpful discussions.
The Authors acknowledge support from European Research Council (ERC) grant number:
101002511/project acronym: VEGA P.
{\em Software}: 
\texttt{Numpy} \citep{Harris2020}; 
\texttt{Matplotlib} \citep{Hunter2007}; 
\texttt{SciPy} \citep{Virtanen2020};
\texttt{Astropy} \citep{Astropy13,Astropy18,Astropy22};

\vspace{-0.8em}

\vspace{-1.0em}
\bibliographystyle{mnras}
\bibliography{bibliography}

\begin{thebibliography}{}
\makeatletter
\relax
\def\mn@urlcharsother{\let\do\@makeother \do\$\do\&\do\#\do\^\do\_\do\%\do\~}
\def\mn@doi{\begingroup\mn@urlcharsother \@ifnextchar [ {\mn@doi@} {\mn@doi@[]}}
\def\mn@doi@[#1]#2{\def\@tempa{#1}\ifx\@tempa\@empty \href {http://dx.doi.org/#2} {doi:#2}\else \href {http://dx.doi.org/#2} {#1}\fi \endgroup}
\def\mn@eprint#1#2{\mn@eprint@#1:#2::\@nil}
\def\mn@eprint@arXiv#1{\href {http://arxiv.org/abs/#1} {{\tt arXiv:#1}}}
\def\mn@eprint@dblp#1{\href {http://dblp.uni-trier.de/rec/bibtex/#1.xml} {dblp:#1}}
\def\mn@eprint@#1:#2:#3:#4\@nil{\def\@tempa {#1}\def\@tempb {#2}\def\@tempc {#3}\ifx \@tempc \@empty \let \@tempc \@tempb \let \@tempb \@tempa \fi \ifx \@tempb \@empty \def\@tempb {arXiv}\fi \@ifundefined {mn@eprint@\@tempb}{\@tempb:\@tempc}{\expandafter \expandafter \csname mn@eprint@\@tempb\endcsname \expandafter{\@tempc}}}

\bibitem[\protect\citeauthoryear{{Antonini}, {Faber}, {Gualandris}  \& {Merritt}}{{Antonini} et~al.}{2010}]{AntoniniFaber2010}
{Antonini} F.,  {Faber} J.,  {Gualandris} A.,   {Merritt} D.,  2010, \mn@doi [\apj] {10.1088/0004-637X/713/1/90}, \href {https://ui.adsabs.harvard.edu/abs/2010ApJ...713...90A} {713, 90}

\bibitem[\protect\citeauthoryear{{Antonini}, {Lombardi}  \& {Merritt}}{{Antonini} et~al.}{2011}]{AntoniniLombardi2011}
{Antonini} F.,  {Lombardi} Jr. J.~C.,   {Merritt} D.,  2011, \mn@doi [\apj] {10.1088/0004-637X/731/2/128}, \href {https://ui.adsabs.harvard.edu/abs/2011ApJ...731..128A} {731, 128}

\bibitem[\protect\citeauthoryear{{Astropy Collaboration} et~al.,}{{Astropy Collaboration} et~al.}{2013}]{Astropy13}
{Astropy Collaboration} et~al., 2013, \mn@doi [\aap] {10.1051/0004-6361/201322068}, \href {https://ui.adsabs.harvard.edu/abs/2013A&A...558A..33A} {558, A33}

\bibitem[\protect\citeauthoryear{{Astropy Collaboration} et~al.,}{{Astropy Collaboration} et~al.}{2018}]{Astropy18}
{Astropy Collaboration} et~al., 2018, \mn@doi [\aj] {10.3847/1538-3881/aabc4f}, \href {https://ui.adsabs.harvard.edu/abs/2018AJ....156..123A} {156, 123}

\bibitem[\protect\citeauthoryear{{Astropy Collaboration} et~al.,}{{Astropy Collaboration} et~al.}{2022}]{Astropy22}
{Astropy Collaboration} et~al., 2022, \mn@doi [\apj] {10.3847/1538-4357/ac7c74}, \href {https://ui.adsabs.harvard.edu/abs/2022ApJ...935..167A} {935, 167}

\bibitem[\protect\citeauthoryear{{Becklin} \& {Neugebauer}}{{Becklin} \& {Neugebauer}}{1968}]{Becklin}
{Becklin} E.~E.,  {Neugebauer} G.,  1968, \mn@doi [\apj] {10.1086/149425}, \href {https://ui.adsabs.harvard.edu/abs/1968ApJ...151..145B} {151, 145}

\bibitem[\protect\citeauthoryear{{Bradnick}, {Mandel}  \& {Levin}}{{Bradnick} et~al.}{2017}]{BradnickMandelLevin2017}
{Bradnick} B.,  {Mandel} I.,   {Levin} Y.,  2017, \mn@doi [\mnras] {10.1093/mnras/stx1007}, \href {https://ui.adsabs.harvard.edu/abs/2017MNRAS.469.2042B} {469, 2042}

\bibitem[\protect\citeauthoryear{{Bromley}, {Kenyon}, {Geller}, {Barcikowski}, {Brown}  \& {Kurtz}}{{Bromley} et~al.}{2006}]{BromleyKenion2006}
{Bromley} B.~C.,  {Kenyon} S.~J.,  {Geller} M.~J.,  {Barcikowski} E.,  {Brown} W.~R.,   {Kurtz} M.~J.,  2006, \mn@doi [\apj] {10.1086/508419}, \href {https://ui.adsabs.harvard.edu/abs/2006ApJ...653.1194B} {653, 1194}

\bibitem[\protect\citeauthoryear{{Bromley}, {Kenyon}, {Brown}  \& {Geller}}{{Bromley} et~al.}{2018}]{BromleyKenyon18}
{Bromley} B.~C.,  {Kenyon} S.~J.,  {Brown} W.~R.,   {Geller} M.~J.,  2018, \mn@doi [\apj] {10.3847/1538-4357/aae83e}, \href {https://ui.adsabs.harvard.edu/abs/2018ApJ...868...25B} {868, 25}

\bibitem[\protect\citeauthoryear{{Brown}, {Geller}  \& {Kenyon}}{{Brown} et~al.}{2014}]{Brown2014}
{Brown} W.~R.,  {Geller} M.~J.,   {Kenyon} S.~J.,  2014, \mn@doi [\apj] {10.1088/0004-637X/787/1/89}, \href {https://ui.adsabs.harvard.edu/abs/2014ApJ...787...89B} {787, 89}

\bibitem[\protect\citeauthoryear{{Brown}, {Kobayashi}, {Rossi}  \& Re'em}{{Brown} et~al.}{2018a}]{Brown18}
{Brown} H.,  {Kobayashi} S.,  {Rossi} E.~M.,   Re'em S.,  2018a, \mn@doi [\mnras] {10.1093/mnras/sty1069}, 477, 5682–5691

\bibitem[\protect\citeauthoryear{{Brown}, {Lattanzi}, {Kenyon}  \& {Geller}}{{Brown} et~al.}{2018b}]{BrownLattanzi18}
{Brown} W.~R.,  {Lattanzi} M.~G.,  {Kenyon} S.~J.,   {Geller} M.~J.,  2018b, \mn@doi [\apj] {10.3847/1538-4357/aadb8e}, \href {https://ui.adsabs.harvard.edu/abs/2018ApJ...866...39B} {866, 39}

\bibitem[\protect\citeauthoryear{{Campbell}, {Ciurlo}  \& {Morris}}{{Campbell} et~al.}{2023}]{Campbell}
{Campbell} R.,  {Ciurlo} A.,   {Morris} M.,  2023, in American Astronomical Society Meeting Abstracts. p. 311.01

\bibitem[\protect\citeauthoryear{{Capuzzo-Dolcetta} \& {Fragione}}{{Capuzzo-Dolcetta} \& {Fragione}}{2015}]{Capuzzo2015}
{Capuzzo-Dolcetta} R.,  {Fragione} G.,  2015, \mn@doi [\mnras] {10.1093/mnras/stv2123}, \href {https://ui.adsabs.harvard.edu/abs/2015MNRAS.454.2677C} {454, 2677}

\bibitem[\protect\citeauthoryear{{Chu} et~al.,}{{Chu} et~al.}{2023}]{chu}
{Chu} D.~S.,  et~al., 2023, \mn@doi [\apj] {10.3847/1538-4357/acc93e}, \href {https://ui.adsabs.harvard.edu/abs/2023ApJ...948...94C} {948, 94}

\bibitem[\protect\citeauthoryear{{Ciurlo} et~al.,}{{Ciurlo} et~al.}{2020}]{CiurloNaoz2020}
{Ciurlo} A.,  et~al., 2020, \mn@doi [\nat] {10.1038/s41586-019-1883-y}, \href {https://ui.adsabs.harvard.edu/abs/2020Natur.577..337C} {577, 337}

\bibitem[\protect\citeauthoryear{{Ciurlo} et~al.,}{{Ciurlo} et~al.}{2021}]{Ciurlo}
{Ciurlo} A.,  et~al., 2021, in {Tsuboi} M.,  {Oka} T.,  eds,  Astronomical Society of the Pacific Conference Series Vol. 528, New Horizons in Galactic Center Astronomy and Beyond. p.~215

\bibitem[\protect\citeauthoryear{{Eggleton}}{{Eggleton}}{1983}]{Eggleton1983}
{Eggleton} P.~P.,  1983, \mn@doi [\apj] {10.1086/160960}, \href {https://ui.adsabs.harvard.edu/abs/1983ApJ...268..368E} {268, 368}

\bibitem[\protect\citeauthoryear{{Evans}, {Marchetti}  \& {Rossi}}{{Evans} et~al.}{2022a}]{Evans2022}
{Evans} F.~A.,  {Marchetti} T.,   {Rossi} E.~M.,  2022a, \mn@doi [\mnras] {10.1093/mnras/stac495}, \href {https://ui.adsabs.harvard.edu/abs/2022MNRAS.512.2350E} {512, 2350}

\bibitem[\protect\citeauthoryear{{Evans}, {Marchetti}  \& {Rossi}}{{Evans} et~al.}{2022b}]{Evans2022II}
{Evans} F.~A.,  {Marchetti} T.,   {Rossi} E.~M.,  2022b, \mn@doi [\mnras] {10.1093/mnras/stac2865}, \href {https://ui.adsabs.harvard.edu/abs/2022MNRAS.517.3469E} {517, 3469}

\bibitem[\protect\citeauthoryear{{Evans}, {Rasskazov}, {Remmelzwaal}, {Marchetti}, {Castro-Ginard}, {Rossi}  \& {Bovy}}{{Evans} et~al.}{2023}]{Evans2023}
{Evans} F.~A.,  {Rasskazov} A.,  {Remmelzwaal} A.,  {Marchetti} T.,  {Castro-Ginard} A.,  {Rossi} E.~M.,   {Bovy} J.,  2023, \mn@doi [\mnras] {10.1093/mnras/stad2273}, \href {https://ui.adsabs.harvard.edu/abs/2023MNRAS.525..561E} {525, 561}

\bibitem[\protect\citeauthoryear{{Fragione} \& {Capuzzo-Dolcetta}}{{Fragione} \& {Capuzzo-Dolcetta}}{2016}]{Fragione2016}
{Fragione} G.,  {Capuzzo-Dolcetta} R.,  2016, \mn@doi [\mnras] {10.1093/mnras/stw531}, \href {https://ui.adsabs.harvard.edu/abs/2016MNRAS.458.2596F} {458, 2596}

\bibitem[\protect\citeauthoryear{{Fragione}, {Capuzzo-Dolcetta}  \& {Kroupa}}{{Fragione} et~al.}{2017}]{Fragione2017}
{Fragione} G.,  {Capuzzo-Dolcetta} R.,   {Kroupa} P.,  2017, \mn@doi [\mnras] {10.1093/mnras/stx106}, \href {https://ui.adsabs.harvard.edu/abs/2017MNRAS.467..451F} {467, 451}

\bibitem[\protect\citeauthoryear{{Generozov} \& {Perets}}{{Generozov} \& {Perets}}{2022}]{GenerozovPerets2022}
{Generozov} A.,  {Perets} H.~B.,  2022, \mn@doi [\mnras] {10.1093/mnras/stac1108}, \href {https://ui.adsabs.harvard.edu/abs/2022MNRAS.513.4257G} {513, 4257}

\bibitem[\protect\citeauthoryear{{Ghez} et~al.,}{{Ghez} et~al.}{2003}]{Ghez2003}
{Ghez} A.~M.,  et~al., 2003, \mn@doi [\apjl] {10.1086/374804}, \href {https://ui.adsabs.harvard.edu/abs/2003ApJ...586L.127G} {586, L127}

\bibitem[\protect\citeauthoryear{{Ghez}, {Salimi}, {Hornstein}, A., {Lu}, {Morris}, E.  \& G.}{{Ghez} et~al.}{2005}]{Ghez2005}
{Ghez} A.~M.,  {Salimi} S.,  {Hornstein} S.~D.,  A. T.,  {Lu} J.~R.,  {Morris} M.,  E. B.~E.,   G. D.,  2005, \mn@doi [\apjl] {10.1086/427175}, 620

\bibitem[\protect\citeauthoryear{{Ghez} et~al.,}{{Ghez} et~al.}{2008}]{Ghez2008}
{Ghez} A.~M.,  et~al., 2008, \mn@doi [\apj] {10.1086/592738}, \href {https://ui.adsabs.harvard.edu/abs/2008ApJ...689.1044G} {689, 1044}

\bibitem[\protect\citeauthoryear{{Gillessen}, {Eisenhauer}, {Trippe}, {Alexander}, {Genzel}, {Martins}  \& {Ott}}{{Gillessen} et~al.}{2009}]{Gillessen2009}
{Gillessen} S.,  {Eisenhauer} F.,  {Trippe} S.,  {Alexander} T.,  {Genzel} R.,  {Martins} F.,   {Ott} T.,  2009, \mn@doi [\apj] {10.1088/0004-637X/692/2/1075}, \href {https://ui.adsabs.harvard.edu/abs/2009ApJ...692.1075G} {692, 1075}

\bibitem[\protect\citeauthoryear{{Ginsburg} \& {Loeb}}{{Ginsburg} \& {Loeb}}{2007}]{GinsburgLoeb2007}
{Ginsburg} I.,  {Loeb} A.,  2007, \mn@doi [\mnras] {10.1111/j.1365-2966.2007.11461.x}, \href {https://ui.adsabs.harvard.edu/abs/2007MNRAS.376..492G} {376, 492}

\bibitem[\protect\citeauthoryear{{Harris} et~al.,}{{Harris} et~al.}{2020}]{Harris2020}
{Harris} C.,  et~al., 2020, \nat, 585, 357

\bibitem[\protect\citeauthoryear{{Hills}}{{Hills}}{1988}]{Hills1988}
{Hills} J.~G.,  1988, \mn@doi [\nat] {10.1038/331687a0}, \href {https://ui.adsabs.harvard.edu/abs/1988Natur.331..687H} {331, 687}

\bibitem[\protect\citeauthoryear{{Hunter}}{{Hunter}}{2007}]{Hunter2007}
{Hunter} J.~D.,  2007, \mn@doi [Computing in Science and Engineering] {10.1109/MCSE.2007.55}, \href {https://ui.adsabs.harvard.edu/abs/2007CSE.....9...90H} {9, 90}

\bibitem[\protect\citeauthoryear{{Jansky}}{{Jansky}}{1937}]{Jansky}
{Jansky} K.,  1937, Proceedings of the Institute of Radio Engineers, 25, 1387

\bibitem[\protect\citeauthoryear{{Jia} et~al.,}{{Jia} et~al.}{2023}]{Jia}
{Jia} S.,  et~al., 2023, \mn@doi [\apj] {10.3847/1538-4357/acb939}, \href {https://ui.adsabs.harvard.edu/abs/2023ApJ...949...18J} {949, 18}

\bibitem[\protect\citeauthoryear{{Kenyon}, {Bromley}, {Brown}  \& {Geller}}{{Kenyon} et~al.}{2018}]{KenyonHVSLMC}
{Kenyon} S.~J.,  {Bromley} B.~C.,  {Brown} W.~R.,   {Geller} M.~J.,  2018, \mn@doi [\apj] {10.3847/1538-4357/aada04}, \href {https://ui.adsabs.harvard.edu/abs/2018ApJ...864..130K} {864, 130}

\bibitem[\protect\citeauthoryear{{Kobayashi}, {Hainick}, {Sari}  \& {Rossi}}{{Kobayashi} et~al.}{2012}]{Kobayashi12}
{Kobayashi} S.,  {Hainick} Y.,  {Sari} R.,   {Rossi} E.~M.,  2012, \mn@doi [\apjl] {10.1088/0004-637X/748/2/105}, 748, 105

\bibitem[\protect\citeauthoryear{{Koposov} et~al.,}{{Koposov} et~al.}{2020}]{Koposov}
{Koposov} S.~E.,  et~al., 2020, \mn@doi [\mnras] {10.1093/mnras/stz3081}, 491, 2465–2480

\bibitem[\protect\citeauthoryear{{Landau} \& {Lifshitz}}{{Landau} \& {Lifshitz}}{1976}]{Landau}
{Landau} L.,  {Lifshitz} E.,  1976, { Mechanics}

\bibitem[\protect\citeauthoryear{{Linial} \& {Sari}}{{Linial} \& {Sari}}{2023}]{LinialSari}
{Linial} I.,  {Sari} R.,  2023, \mn@doi [\apj] {10.3847/1538-4357/acbd3d}, \href {https://ui.adsabs.harvard.edu/abs/2023ApJ...945...86L} {945, 86}

\bibitem[\protect\citeauthoryear{{Mandel} \& {Levin}}{{Mandel} \& {Levin}}{2015}]{MandelLevin2015}
{Mandel} I.,  {Levin} Y.,  2015, \mn@doi [\apjl] {10.1088/2041-8205/805/1/L4}, \href {https://ui.adsabs.harvard.edu/abs/2015ApJ...805L...4M} {805, L4}

\bibitem[\protect\citeauthoryear{{Marchetti}, {Contigiani}, {Rossi}, {Albert}, {Brown}  \& {Sesana}}{{Marchetti} et~al.}{2018}]{Marchetti2018}
{Marchetti} T.,  {Contigiani} O.,  {Rossi} E.~M.,  {Albert} J.~G.,  {Brown} A.~G.~A.,   {Sesana} A.,  2018, \mn@doi [\mnras] {10.1093/mnras/sty579}, \href {https://ui.adsabs.harvard.edu/abs/2018MNRAS.476.4697M} {476, 4697}

\bibitem[\protect\citeauthoryear{{Marchetti}, {Evans}  \& {Rossi}}{{Marchetti} et~al.}{2022}]{Marchetti2022}
{Marchetti} T.,  {Evans} F.~A.,   {Rossi} E.~M.,  2022, \mn@doi [\mnras] {10.1093/mnras/stac1777}, \href {https://ui.adsabs.harvard.edu/abs/2022MNRAS.515..767M} {515, 767}

\bibitem[\protect\citeauthoryear{{Markoff} \& {Event Horizon Telescope Collaboration}}{{Markoff} \& {Event Horizon Telescope Collaboration}}{2022}]{EHT2022}
{Markoff} S.,  {Event Horizon Telescope Collaboration} 2022, in American Astronomical Society Meeting \#240. p. 211.01

\bibitem[\protect\citeauthoryear{{McMillan}}{{McMillan}}{2017}]{McMillan2017}
{McMillan} P.~J.,  2017, \mn@doi [\mnras] {10.1093/mnras/stw2759}, \href {https://ui.adsabs.harvard.edu/abs/2017MNRAS.465...76M} {465, 76}

\bibitem[\protect\citeauthoryear{{Miller}, {Freitag}, {Hamilton}  \& {Lauburg}}{{Miller} et~al.}{2005}]{miller05}
{Miller} M.~C.,  {Freitag} M.,  {Hamilton} D.~P.,   {Lauburg} V.~M.,  2005, \mn@doi [\apjl] {10.1086/497335}, \href {https://ui.adsabs.harvard.edu/abs/2005ApJ...631L.117M} {631, L117}

\bibitem[\protect\citeauthoryear{{Murray} \& {Dermott}}{{Murray} \& {Dermott}}{1999}]{Murray}
{Murray} C.~D.,  {Dermott} S.~F.,  1999, {Solar System Dynamics}, \mn@doi{10.1017/CBO9781139174817.
}

\bibitem[\protect\citeauthoryear{Penoyre, Rossi  \& Stone}{Penoyre et~al.}{2025}]{Penoyre2025}
Penoyre Z.,  Rossi E.~M.,   Stone N.~C.,  2025, Disruptions of stars and binary systems on chaotic orbits in the axisymmetric Milky Way center (\mn@eprint {arXiv} {2505.06344}), \url {https://arxiv.org/abs/2505.06344}

\bibitem[\protect\citeauthoryear{{Prodan}, {Antonini}  \& {Perets}}{{Prodan} et~al.}{2015}]{ProdanAntonini2015}
{Prodan} S.,  {Antonini} F.,   {Perets} H.~B.,  2015, \mn@doi [\apj] {10.1088/0004-637X/799/2/118}, \href {https://ui.adsabs.harvard.edu/abs/2015ApJ...799..118P} {799, 118}

\bibitem[\protect\citeauthoryear{{Rasskazov}, {Fragione}, {Leigh}, {Tagawa}, {Sesana}, {Price-Whelan}  \& {Rossi}}{{Rasskazov} et~al.}{2019}]{Rasskazov2019}
{Rasskazov} A.,  {Fragione} G.,  {Leigh} N. W.~C.,  {Tagawa} H.,  {Sesana} A.,  {Price-Whelan} A.,   {Rossi} E.~M.,  2019, \mn@doi [\apj] {10.3847/1538-4357/ab1c5d}, \href {https://ui.adsabs.harvard.edu/abs/2019ApJ...878...17R} {878, 17}

\bibitem[\protect\citeauthoryear{{Raveh} \& {Perets}}{{Raveh} \& {Perets}}{2021}]{raveh21}
{Raveh} Y.,  {Perets} H.~B.,  2021, \mn@doi [\mnras] {10.1093/mnras/staa4001}, \href {https://ui.adsabs.harvard.edu/abs/2021MNRAS.501.5012R} {501, 5012}

\bibitem[\protect\citeauthoryear{{Rein} \& {Liu}}{{Rein} \& {Liu}}{2012}]{ReinLiu2012}
{Rein} H.,  {Liu} S.~F.,  2012, \mn@doi [\aap] {10.1051/0004-6361/201118085}, \href {https://ui.adsabs.harvard.edu/abs/2012A&A...537A.128R} {537, A128}

\bibitem[\protect\citeauthoryear{{Rein} \& {Spiegel}}{{Rein} \& {Spiegel}}{2015}]{ReinSpiegel2015}
{Rein} H.,  {Spiegel} D.~S.,  2015, \mn@doi [\mnras] {10.1093/mnras/stu2164}, \href {https://ui.adsabs.harvard.edu/abs/2015MNRAS.446.1424R} {446, 1424}

\bibitem[\protect\citeauthoryear{{Rossi}, {Kobayashi}  \& {Sari}}{{Rossi} et~al.}{2014}]{Rossi2014}
{Rossi} E.~M.,  {Kobayashi} S.,   {Sari} R.,  2014, \mn@doi [\apj] {10.1088/0004-637X/795/2/125}, \href {https://ui.adsabs.harvard.edu/abs/2014ApJ...795..125R} {795, 125}

\bibitem[\protect\citeauthoryear{{Sari} \& {Fragione}}{{Sari} \& {Fragione}}{2019}]{SariFragione}
{Sari} R.,  {Fragione} G.,  2019, \mn@doi [\apj] {10.3847/1538-4357/ab43df}, \href {https://ui.adsabs.harvard.edu/abs/2019ApJ...885...24S} {885, 24}

\bibitem[\protect\citeauthoryear{{Sari}, {Kobayashi}  \& {Rossi}}{{Sari} et~al.}{2010}]{Sari10}
{Sari} R.,  {Kobayashi} S.,   {Rossi} E.~M.,  2010, \mn@doi [\apjl] {10.1088/0004-637X/708/1/605}, 708, 605

\bibitem[\protect\citeauthoryear{{Sch{\"o}del} et~al.,}{{Sch{\"o}del} et~al.}{2002}]{Schodel}
{Sch{\"o}del} R.,  et~al., 2002, \mn@doi [\nat] {10.1038/nature01121}, \href {https://ui.adsabs.harvard.edu/abs/2002Natur.419..694S} {419, 694}

\bibitem[\protect\citeauthoryear{{Sesana}, {Haardt}  \& {Madau}}{{Sesana} et~al.}{2007}]{Sesana2007}
{Sesana} A.,  {Haardt} F.,   {Madau} P.,  2007, \mn@doi [\mnras] {10.1111/j.1745-3933.2007.00331.x}, 379, L45

\bibitem[\protect\citeauthoryear{{Sesana}, {Haardt}  \& {Madau}}{{Sesana} et~al.}{2008}]{Sesana2008}
{Sesana} A.,  {Haardt} F.,   {Madau} P.,  2008, \mn@doi [\apj] {10.1086/590651}, \href {https://ui.adsabs.harvard.edu/abs/2008ApJ...686..432S} {686, 432}

\bibitem[\protect\citeauthoryear{{Stephan}, {Naoz}, {Ghez}, {Witzel}, {Sitarski}, {Do}  \& {Kocsis}}{{Stephan} et~al.}{2016}]{StephanNaoz2016}
{Stephan} A.~P.,  {Naoz} S.,  {Ghez} A.~M.,  {Witzel} G.,  {Sitarski} B.~N.,  {Do} T.,   {Kocsis} B.,  2016, \mn@doi [\mnras] {10.1093/mnras/stw1220}, \href {https://ui.adsabs.harvard.edu/abs/2016MNRAS.460.3494S} {460, 3494}

\bibitem[\protect\citeauthoryear{{Stephan} et~al.,}{{Stephan} et~al.}{2019}]{StephanNaoz}
{Stephan} A.~P.,  et~al., 2019, \mn@doi [\apj] {10.3847/1538-4357/ab1e4d}, \href {https://ui.adsabs.harvard.edu/abs/2019ApJ...878...58S} {878, 58}

\bibitem[\protect\citeauthoryear{{Stone} \& {Metzger}}{{Stone} \& {Metzger}}{2015}]{StoneMetzger}
{Stone} N.,  {Metzger} B.~D.,  2015, in American Astronomical Society Meeting Abstracts \#225. p. 221.05

\bibitem[\protect\citeauthoryear{{Verberne} et~al.,}{{Verberne} et~al.}{2024}]{Sill2024}
{Verberne} S.,  et~al., 2024, \mn@doi [arXiv e-prints] {10.48550/arXiv.2406.14134}, \href {https://ui.adsabs.harvard.edu/abs/2024arXiv240614134V} {p. arXiv:2406.14134}

\bibitem[\protect\citeauthoryear{{Verberne}, {Rossi}, {Koposov}, {Penoyre}, {Cavieres}  \& {Kuijken}}{{Verberne} et~al.}{2025}]{Verberne2025}
{Verberne} S.,  {Rossi} E.~M.,  {Koposov} S.~E.,  {Penoyre} Z.,  {Cavieres} M.,   {Kuijken} K.,  2025, \mn@doi [arXiv e-prints] {10.48550/arXiv.2502.17165}, \href {https://ui.adsabs.harvard.edu/abs/2025arXiv250217165V} {p. arXiv:2502.17165}

\bibitem[\protect\citeauthoryear{{Virtanen} et~al.,}{{Virtanen} et~al.}{2020}]{Virtanen2020}
{Virtanen} P.,  et~al., 2020, \nat, 17, 261

\bibitem[\protect\citeauthoryear{{Yu} \& {Tremaine}}{{Yu} \& {Tremaine}}{2003}]{Yu2003}
{Yu} Q.,  {Tremaine} S.,  2003, \mn@doi [\apj] {10.1086/379546}, \href {https://ui.adsabs.harvard.edu/abs/2003ApJ...599.1129Y} {599, 1129}

\makeatother
\end{thebibliography}


\appendix

\section{Effect of eccentricity}
\label{app:ecc}
\begin{figure}
    \centering
    \includegraphics[width=0.85\columnwidth]{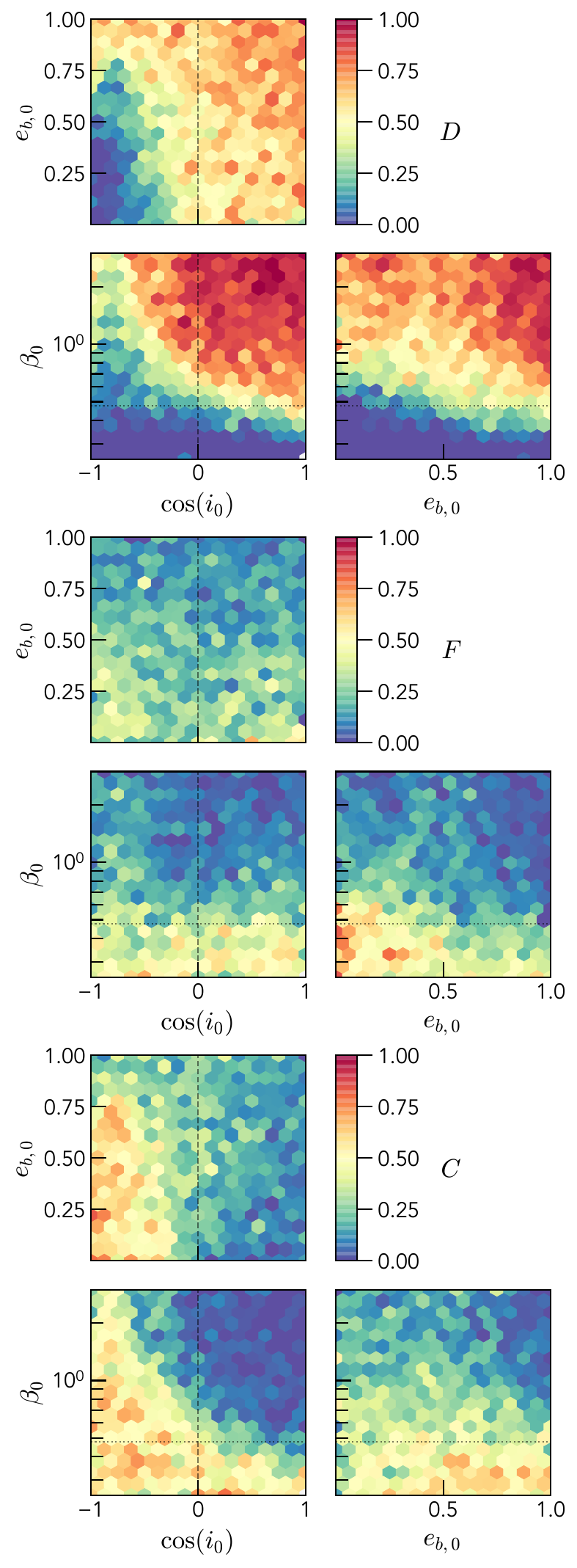}
    \caption{Fractions of disruptions (D, top), fly-aways (F, middle) and coming-backs (C, bottom) after a single passage for binaries with a flat distribution of initial eccentricities. We show the outcomes as a function of the initial conditions $\beta_0$, $\cos(i_0)$ and $e_{\mathrm{b},0}$. For each two dimensional plot the results are marginalized over the third dimension. 
    The vertical dashed line corresponds to an inclination of $\frac{\pi}{2}$:  we call binaries around this angle  \emph{intermediate},  those on its left {\it retrograde}, and those on its right, {\it prograde}. The dotted horizontal line is at $\beta_{\text{lim}}$.
    This figure is generated for 100,000 sets of initial conditions.
    }
    \label{hexbinHCF_ecc}
\end{figure}

Although we start with initially circular binaries those that comeback for subsequent passages can be (and generally are) eccentric. Thus in Fig. \ref{hexbinHCF_ecc} we show the fractions of outcomes (D, F and C) for a flat distribution of eccentricities (as can be compared to Fig. \ref{fig:hexbinHCF}).

We see that more eccentric systems are more susceptible to disruption and allow disruptions at lower $\beta_0$. For high $e_{\mathrm{b},0}$ ($\gtrsim 0.8$) the dependence on inclination almost disappears, with retrograde systems similarly likely to disrupt as prograde. The fraction of FAs is much reduced for even mild eccentricities ($e_{\mathrm{b},0} \gtrsim 0.1$) and correspondingly there are more systems that comeback.

So for initially higher eccentricities, systems are generally more prone to disrupt and CBs more likely (the FAs fraction changes accordingly), with inclination playing a reduced role at high eccentricities.
These results motivates the analysis of multiple passages and help to interpret the consequent results (e.g. the fractions distributions in the lower panel of Fig. \ref{fig:hexbinHCF} or the disruption boost observed in Fig. \ref{fig:allfrac}  after subsequent passages).

\section{Characteristic units of the Hills mechanism}
\label{app:units}

Let us start by considering a generic situation where the energy of a particle of mass $m_i$ is perturbed by the presence of a mass $M$ by an amount $\Delta E$ and its angular momentum by $\Delta \mathbf{L}$; the particle's final properties can be described as
\begin{align}
    E&=E_0+\Delta E,\\
    \mathbf{L}&=\mathbf{L_0}+\Delta \mathbf{L}.
\end{align}
From these, we can define the (change of) characteristic scale of the other orbital properties (i.e. semi-major axis, velocity, eccentricity and inclination, respectively) as
\begin{align}
a&= -\frac{GMm_i}{2(E_0 +\Delta E)}\\
v&=\sqrt{\frac{2 (E_0+\Delta E)}{m_i}}\\
e^2 &= 1 -\frac{2E (\mathbf{L}\cdot \mathbf{L})}{G^2 M^2 m_i^3} = 1-\frac{2E_0 L_0^2}{G^2 M^2 m_i^3}  \\
&- \frac{2}{G^2 M^2 m_i^3}\left((E_0+\Delta E) (2 (\mathbf{L_0}\cdot \Delta \mathbf{L}) +(\Delta \mathbf{L}\cdot \Delta \mathbf{L})) + \Delta E L_0^2\right)\\
\cos(i)&=\frac{1+(\mathbf{\Delta L}\cdot \mathbf{\hat{L}_0})}{\sqrt{1+2\frac{1}{L_0}(\mathbf{\Delta L}\cdot \mathbf{\hat{L}
_0})+\left( \frac{\Delta L}{L_0} \right)^2}}=\frac{1+\Delta L_z}{\sqrt{1+2\frac{\Delta L_z}{L_0}+\left( \frac{\Delta L}{L_0} \right)^2}}\\
\end{align}
where, in the last equality of the expression for $\cos(i)$ we assumed that $\mathbf{L_0}=L_0 \mathbf{\hat{z}}$ (i.e. that the initial orbit is in the x,y plane).

We can define the characteristic scale of $\Delta E$ and $|\Delta \mathbf{L}|$ as $\epsilon$ and $\Lambda$, which then set the characteristic scale of the (change of) other orbital properties. With relevance to this work \citep[following][]{Kobayashi12}, we make the simplifying assumptions that 
\begin{itemize}
    \item $|\Delta E| \sim \epsilon \gg |E_0|$ (i.e. the particle's trajectory is close enough to parabolic to ignore its initial energy);
    \item $ |\Delta \mathbf{L}| \sim \Lambda \ll|\mathbf{L_0}|$ (i.e. the initial orbit has significant angular momentum).
\end{itemize}
The former assumption (of a parabolic trajectory) implies that whether the new orbit of the particle is bound or unbound depends only on the sign of $\Delta E$.

If $\Delta E >0$ the particle's will move a trajectory unbound from the massive perturber and will escape to infinity with a characteristic velocity
\begin{equation}
\nu \sim \sqrt{\frac{2\epsilon}{m_i}}.
\end{equation}

If $\Delta E<0$, the particle moves on a bound orbit around the mass M. The characteristic semi-major axis of the new orbit is
\begin{equation}
\alpha \sim \frac{GMm_i}{2\epsilon}.
\end{equation}
and its eccentricity (in the parabolic case, to first order) is
\begin{equation}
e^2 \sim 1 - \frac{2\Delta E L_0^2}{G^2 M^2 m_i^3}.
\end{equation}
We can express this in terms of $\delta$, which corresponds to the characteristic size of $|1-e|$, as follows
\begin{equation}
\delta \sim \frac{L_0^2 \epsilon}{G^2 M^2 m_i^3}.
\end{equation}

In the parabolic approximation and making a small angle approximation as $\sqrt{L_x^2 + L_y^2} = \Delta L_t \ll L_0$, 
we can derive the characteristic change in inclination as
\begin{equation}
\eta \sim \frac{\Lambda_t}{L_0} \sim \frac{\Lambda}{\sqrt{2} L_0}
\end{equation}
where $\Lambda_t$ is the characteristic change in angular momentum tangential to the initial direction. We have found it a reasonable assumption for the Hills mechanism that $\Lambda_t \sim \Lambda_z \sim \frac{\Lambda}{\sqrt{2}}$ (i.e. that on average there is equal transfer of angular momentum in the perpendicular and parallel directions).


\subsection{Characteristic change of energy and angular momentum}

We now determine the values for $\epsilon$ and $\Lambda$ corresponding to the dimensions of the problem at hand (for a bound binary and a disrupted one, respectively).

\subsubsection{Characteristics of a bound binary}

We now consider an initial binary with masses $m_1$ and $m_2$, mass ratio $q=m_2/m_1<1$ and initial semi-major axis $a_{\mathrm{b},i}$, moving around and MBH with mass $M$ (with $Q=M/m\gg1$). Its initial energy is given by
\begin{equation}
E_{\mathrm{b},i} = -\frac{G m_1 m_2}{2 a_{\mathrm{b},i}} = -\frac{G \mu m}{2 a_{\mathrm{b},i}} 
\end{equation}
Any change in energy of the binary, $\Delta E_{\rm b}$, while the binary remains bound, will be of order $|E_{\mathrm{b},i}|$. Thus, it is natural to define a characteristic energy\footnote{As $E_{\rm b}$ is the most energy a binary could gain and remain bound this characteristic energy is an upper limit, and thus all characteristic units that depend on $\epsilon_{\rm B}$ will be upper or lower limits.} of the binary
\begin{equation}
\epsilon_{\rm b} = |E_{\mathrm{b},i}| =  \frac{G \mu m}{2 a_{\mathrm{b},i}}
\end{equation}
where the reduced mass $\mu=m_1 m_2/m$.

Similarly we can define the characteristic angular momentum of the binary in terms of the maximum angular momentum (for a given energy) corresponding to a circular orbit
\begin{equation}
\Lambda_{\rm b} =L_{\mathrm{b},\mathrm{circ},i} = \mu \sqrt{G m a_{\mathrm{b},i}}.
\end{equation}

If we consider the restricted three-body treatment of the problem and use the natural units of the Hills mechanism introduced in section \ref{sec:units}, we can express the above scales as
$\epsilon_{\rm b}=\frac{1}{2\beta}\frac{q}{(1+q)^2}\frac{m\lambda^2}{\tau^2}$ and $\Lambda_{\rm b}=\sqrt{\beta}\frac{q}{(1+q)^2}\frac{m\lambda^2}{\tau}$.

\subsubsection{Characteristics of a disrupted binary}

The characteristic units in case of disruption can be found in the high $\beta$ limit (though, as shown in the main text, they agree within a factor of a few across all $\beta$). Extreme $\beta$ corresponds to an $r_{\rm p} \rightarrow 0$, which reduces to the simpler dynamical case of radial infall of the CM. Taking $t=0$ as the moment when $r_{\mathrm{cm}}=0$, then the CM motion follows
\begin{align}
r_{\mathrm{cm}} &= \left( \frac{9 G M t^2}{2} \right)^\frac{1}{3},\\
v_{\mathrm{cm}} &= \frac{1}{t}\left( \frac{4 G M t^2}{3} \right)^\frac{1}{3}
\end{align}
(with $v_{\mathrm{cm}}$ negative during infall, $t<0$, and positive afterwards). At large times, the distance goes to infinity and the velocity to 0, and thus the energy of the CM-orbit is 0. Similarly, as the motion is along a straight line towards the origin the angular momentum is also zero.

To determine the corresponding characteristic scales, we can approximate the true behavior of the binary assuming that it is unaffected by the MBH until it reaches the tidal radius, and completely dominated by it after that point. This approximation is the more accurate the deeper the encounter is (large $\beta$ limit).
Thus, the binary will have characteristic separation $a_{\rm b}$ and speed $\sqrt{\frac{Gm}{a_{\rm b}}}$ up to and including the moment it reaches the tidal radius.

At the moment of separation ($r_{\mathrm{cm}}=r_{\rm t}$), $t=t_t = -\sqrt{\frac{2 r_{\rm t}^3}{9 GM}}$ and $v_{\mathrm{cm}}=v_t=-\sqrt{\frac{2GM}{r_{\rm t}}}$. Taking the binary's instantaneous relative displacement and velocity to be $\mathbf{r}=\mathbf{r}_2 - \mathbf{r}_1$ and $\mathbf{v}=\mathbf{v}_2 - \mathbf{v}_1$, respectively, then the positions and velocities of either mass 1 or 2 is
\begin{equation}
\mathbf{r_{12}} = \mathbf{r_{\mathrm{cm}}} \mp \frac{m_{21}}{m} \mathbf{r}  \ \ \ \mathrm{and} \ \ \ \mathbf{v_{12}} = \mathbf{v_{\mathrm{cm}}} \mp \frac{m_{21}}{m} \mathbf{v}.
\end{equation}

Until the binary has separated we have that $|\mathbf{r}| \ll |\mathbf{r}_{\mathrm{cm}}|$ and $|\mathbf{v}| \ll |\mathbf{v}_{\mathrm{cm}}|$; thus, we can expand the energy and angular momentum to first order as
\begin{equation}
\begin{aligned}
E_{1,2} &= \frac{m (\mathbf{v_{12}}\cdot\mathbf{v_{12}})}{2}- \frac{G M m}{|\mathbf{r_{12}}|} \\
&\sim \frac{m_{12}}{m} E_0 \mp \frac{m_1 m_2}{m}\left( \mathbf{v}_{\mathrm{cm}} \cdot \mathbf{v} + \frac{GM}{r_{\mathrm{cm}}^3}(\mathbf{r}_{\mathrm{cm}} \cdot \mathbf{r})\right)
\end{aligned}
\end{equation}
and
\begin{equation}
\begin{aligned}
\mathbf{L}_{1,2} &= m_{12} \mathbf{r}_{12} \wedge \mathbf{v}_{12} \\
&\sim \frac{m_{12}}{m} \mathbf{L}_0 \mp \frac{m_1 m_2}{m}\left(\mathbf{r} \wedge \mathbf{v}_{{\mathrm{cm}}} + \mathbf{r}_{\mathrm{cm}} \wedge \mathbf{v} \right),
\end{aligned}
\end{equation}
where $E_0$ and $\mathbf{L}_0$ are the initial energy and angular momentum of the CM.

We now consider our case of interest, where $E_0$ and $L_0 = 0$ (radial case). We substitute for the centre of mass $|\mathbf{r}_{\mathrm{cm}}|\sim r_{\rm t}$, $|\mathbf{v}_{\mathrm{cm}}|\sim v_t$ and for the binary $|\mathbf{r}|\sim a_{\rm b}$, $|\mathbf{v}|\sim \sqrt{\frac{Gm}{a_{\rm b}}}$ and use $r_{\rm t} \sim Q^\frac{1}{3} a_{\rm b}$. Then, ignoring geometric terms of order unity, the characteristic changes in energy and angular momentum are, respectively:
\begin{align}
\epsilon_{\rm D} &= Q^\frac{1}{3} \frac{G m_1 m_2}{a_{\rm b}} = 2 Q^\frac{1}{3} \epsilon_{\rm b},\\
\Lambda_{\rm D} &= Q^\frac{1}{3} \frac{m_1 m_2}{m} \sqrt{G m a_{\rm b}} = Q^\frac{1}{3} \Lambda_{\rm b}.
\end{align}

In characteristic units they can be expressed as $\epsilon_{\rm D} = \frac{Q^\frac{1}{3}}{\beta} \frac{q}{(1+q)^2}\frac{m\lambda^2}{\tau^2}$ and $\Lambda_{\rm D} = Q^\frac{1}{3}\sqrt{\beta}\frac{q}{(1+q)^2}\frac{m\lambda^2}{\tau}$.

These characteristic units hold true for a wide range of $\beta$ ($\gtrsim 1$) where the assumption of a radial orbit with zero angular momentum is no longer true. For general $\beta$ the initial angular momentum of the centre of mass orbit is
\begin{equation}
L_{\mathrm{cm}}=\sqrt{(1+e_{\mathrm{cm}})GM m^2 r_{\rm p}} = \sqrt{\frac{ a_{\rm b}}{\beta}}\sqrt{2 G Q^\frac{4}{3} m^3}.
\end{equation}

\subsection{Characteristic scales of orbital properties}
From the characteristic energy and angular momentum scales obtained in the previous section, we now derive the corresponding scales for the orbital properties of bound and disrupted binaries. 
\subsubsection{Bound binaries}
For surviving binaries (e.g. FAs), $m_i=m$ and $L_0 =L_{\rm cm}$, and the relevant characteristic units to use are $\epsilon_{\rm B}$ and $\Lambda_{\rm B}$.
These translate to the following characteristic scales for semi-major axis, velocity, deviation of eccentricity from 1 and inclination, respectively: 
\begin{align}
\alpha_{\rm b}&= Q\frac{m}{\mu} a_{\mathrm{b}},\\
\nu_{\rm b}&= \sqrt{\frac{G\mu}{a_{\mathrm{b}}}},\\
\delta_{\rm b}&= Q^{-\frac{2}{3}}\frac{1}{\beta} \frac{\mu}{m},\\
\eta_{\rm b}&= Q^{-\frac{1}{2}}\frac{\sqrt{ \beta}}{2} \frac{\mu}{m}.
\end{align}

\subsubsection{Disrupted binaries}
For disrupted binaries the characteristic units are $\epsilon_{\rm D}$ and $\lambda_{\rm D}$ and the mass of interest is $m_{\rm i}$ is either $m_1$ or $m_2$. Now $L_0=\frac{m_{\rm i}}{m}L_{\mathrm{cm}}$ (and $E_0=0$). If we define the factor
\begin{equation}
\Gamma \equiv 2 Q^\frac{1}{3} \frac{m}{m_{\rm i}}.
\end{equation}
then the characteristic scales for a disruption can be written simply as:
\begin{align}
\nu_{\rm D}&= \Gamma^\frac{1}{2} \cdot \nu_{\rm b},\\
\alpha_{\rm D}&= \Gamma^{-1} \cdot \alpha_{\rm b},\\
\delta_{\rm D}&= \Gamma \cdot \delta_{\rm b},\\
\eta_{\rm D}&= \frac{1}{2}\Gamma \cdot \eta_{\rm b}.
\end{align} 

Given that $\Gamma$ is significantly greater than one, we can see that disrupted binaries result in faster ejections, much more eccentric and inclined orbits, and substantially tighter orbits with respect to bound binaries.

We note that while $\Gamma$ can be arbitrarily large for a small $m_2$, it is, in every case, balanced by $\mu \rightarrow m_2$. In these cases, namely when $q \ll 1$, the properties of surviving binaries are barely changed, and its only the lighter companion that can have extreme velocities.

Re-expressing these in terms of the physical scales of the problem we obtain,
\begin{equation}
    \nu_{\rm D}=Q^{\frac{1}{6}}\sqrt{2G\frac{m \mu}{m_{\rm i}a_{\mathrm{b}}}},
\end{equation}
\begin{align}
    \alpha_{\rm D}&=\frac{1}{2}Q^{\frac{2}{3}}\frac{m_{\rm i}}{\mu}a_{\mathrm{b}},
\end{align}
\begin{equation}
\delta_{\rm D}=2Q^{-\frac{1}{3}}\beta^{-1}\frac{\mu}{m_{\rm i}}.
\end{equation}
and
\begin{equation}
\eta_{\rm D}=\frac{1}{2}Q^{-\frac{1}{6}}\beta^{\frac{1}{2}}\frac{\mu}{m_{\rm i}},
\end{equation}

\bsp
\label{lastpage}
\end{document}